\newif\ifdraft\drafttrue
\newif\ifcolor\colortrue
\newif\iftechreport\techreporttrue

\iftechreport
\documentclass[onecolumn]{article}
\pdfoutput=1
\usepackage{epsfig}
\usepackage{amssymb}
\usepackage{amsmath}
\usepackage{amsfonts}
\usepackage[margin=1.0in]{geometry}
\else
\documentclass[letter]{sig-alternate-10pt}
\fi

\usepackage[table]{xcolor}
\definecolor{pennblue}{cmyk}{1,.65,0,.3}
\definecolor{pennred}{cmyk}{0,1,.65,.34}

\definecolor{williamspurple}{RGB}{89,23,128}
\definecolor{williamsgold}{RGB}{255,213,0}

\usepackage{arydshln}
\usepackage{booktabs}
\usepackage{cmtt}

\usepackage{paralist}
\usepackage{lastpage}
\usepackage{times}
\usepackage{tikz}
\usepackage{alltt}
\usepackage{graphicx}
\usepackage{balance}
\usepackage{amsmath}
\usepackage{caption}
\usepackage{subcaption}
\usepackage{algorithm,algpseudocode}
\usepackage{enumitem}

\usepackage{verbatim}

\newcommand{\ax}{4}
\newcommand{\bx}{2.5}
\newcommand{\cx}{4}
\newcommand{\dx}{5.5}
\newcommand{\ex}{5.5}
\newcommand{\fx}{7}
\newcommand{\gx}{8.5}
\newcommand{\hx}{10}
\newcommand{\ix}{8.5}
\newcommand{\tcx}{4.5}
\newcommand{\tex}{5.5}
\newcommand{\tix}{6.8}
\newcommand{\ay}{1.8}
\newcommand{\by}{3}
\newcommand{\cy}{4.2}
\newcommand{\dy}{3}
\newcommand{\ey}{5.4}
\newcommand{\fy}{4.2}
\newcommand{\gy}{5.4}
\newcommand{\hy}{4.2}
\newcommand{\iy}{3}
\newcommand{\tcy}{6.7}
\newcommand{\tey}{8}
\newcommand{\tiy}{6.7}
\newcommand{\shiftamount}{9.3}
\newcommand{\treeedge}{dotted}
\usepackage{tikz}
\pgfdeclarelayer{background}
\pgfdeclarelayer{foreground}
\pgfsetlayers{foreground,background,main}

\ifcolor
\usepackage[colorlinks=true,linkcolor={pennblue},citecolor={pennred},urlcolor={pennblue}]{hyperref}
\else
\usepackage[colorlinks=true,linkcolor={black},citecolor={black},urlcolor={black}]{hyperref}
\fi
\usepackage{url}

\iftechreport
\else
\fi
\usepackage[export]{adjustbox} 

\begin{document}

\iftechreport
\title{Kulfi: Robust Traffic Engineering\\Using Semi-Oblivious Routing}
\fi

\iftechreport
\author{ Praveen Kumar \\ Cornell University
\and Yang Yuan \\ Cornell University
\and Chris Yu \\ Carnegie Mellon University
\and Nate Foster \\ Cornell University
\and Robert Kleinberg \\ Cornell University 
\and Robert Soul\'{e} \\ Universit\`{a} della Svizzera italiana }

\fi

\iftechreport
\date{}
\fi

\maketitle

\begin{abstract}
\iftechreport
Wide-area network traffic engineering enables network operators to reduce congestion and
\fi
improve utilization by balancing load across multiple paths. Current
approaches to traffic engineering can be modeled in terms of a routing
component that computes forwarding paths, and a load balancing
component that maps incoming flows onto those paths dynamically,
adjusting sending rates to fit current conditions.  Unfortunately,
existing systems rely on simple strategies for one or both of these
components, which leads to poor performance or requires making
frequent updates to forwarding paths, significantly increasing
management complexity. This paper explores a different approach based
on \emph{semi-oblivious routing}, a natural extension
of \emph{oblivious routing} in which the system computes a diverse set
of paths independent of demands, but also dynamically adapts sending
rates as conditions change. Semi-oblivious routing has a number of
important advantages over competing approaches including low overhead,
nearly optimal performance, and built-in protection against unexpected
bursts of traffic and failures. Through in-depth simulations and a
deployment on SDN hardware, we show that these benefits are robust,
and hold across a wide range of topologies, demands, resource budgets,
and failure scenarios.

\end{abstract}

\newcommand{\jnfpar}[1]{\ \\[-.75em] \textit{#1.}~}

\section{Introduction}
\label{sec:introduction}

%
%
Wide-area network (WAN) traffic engineering (TE) is a topic of interest for many
leading technology companies including Amazon, Facebook, Google, Microsoft,
Netflix, and others. The private networks operated by these companies must
balance demands between latency-sensitive, customer-facing traffic and
high-volume, operational traffic, such as bulk replication for large data
stores.  A common objective is to improve network utilization by distributing
load across multiple paths to obtain better operational efficiency. Towards this
goal, a variety of approaches have been explored in recent years, each relying
on a different combination of underlying strategies for computing paths and
distributing load~\cite{shen08,suchara11,hong13,jain13}.

%
%
The textbook approach to traffic engineering frames it as a
combinatorial optimization problem: given a capacitated network and a
set of demands for flow between sources and destinations, find an
assignment of flows to paths that optimizes for some criterion, such
as minimizing the maximum amount of congestion on any link. This is
known as the \emph{multi-commodity flow} (MCF) problem in the
literature, and has been extensively studied. If the flow between each
source and destination is restricted to use a single path, then the
problem is NP-complete. But, if fractional flows are allowed, then
optimal solutions can be found in polynomial time using linear
programming (LP). Scalability and running time can be further improved
by relaxing the optimality requirement and using an approximation
algorithm, such as the \emph{multiplicative weights}
method~\cite{plotkin95,garg2007faster,AHK12}.

%
%
\jnfpar{Optimal traffic engineering}
In principle, it is possible to build an optimal traffic engineering
system by repeatedly executing the following steps in a loop: (i)
monitor the network load and build up an estimate of the demands for
traffic between each source-destination pair; (ii) translate the
demands into an LP and use an off-the-shelf solver to compute an
optimal set of forwarding paths; and (iii) update the configurations
of network switches to implement those paths. However, to actually
build such a system, one would have to overcome a number of major
practical challenges:
\begin{itemize}[nolistsep,topsep=0em,leftmargin=*]
\item Solving MCF instances could become a bottleneck,
  especially at scale and in dynamic environments where solutions
  would need to be computed frequently to stay close to the optimum.
\item Estimating demands accurately could be difficult, especially
  with unexpected failures and traffic bursts.
\item Switches have limited amounts of memory for implementing paths,
  but it is not clear how to limit MCF to abide by resource budgets.
\item Small changes in demands could produce dramatically 
  different paths, which would lead to excessive churn.
\item Updating configurations in a consistent manner could  be
  difficult and would add significant complexity to the management
  infrastructure~\cite{reitblatt12, jin14}.
\item Switches could take several seconds to
  implement updates, which imposes a limit on how rapidly the system
  can adapt to changing conditions.
\end{itemize}

\jnfpar{Centralized traffic engineering}
One way to side-step these challenges is to pre-compute a set of
forwarding paths and distribute flows onto those paths
dynamically---e.g., using the global visibility offered by a
logically-centralized SDN controller. This approach gives up on
optimality, since it pre-commits to using a restricted set of paths,
but it retains many of the benefits of MCF-based approaches and is
much simpler to implement. In particular, it only requires the ability
to update sending rates at the edge of the network rather than
changing end-to-end forwarding paths at each iteration. By restricting
attention to relatively small topologies, scheduling traffic with
elastic demands, and using heuristic approximation algorithms, systems
such as B4~\cite{jain13} and SWAN~\cite{hong13} have been able to
obtain dramatic improvements in production settings. However, one
important aspect of these systems has been under-explored: the
algorithms they use to select forwarding paths. B4 uses a greedy
heuristic that attempts to ensure fairness while SWAN uses
$k$-shortest paths. Relying on simple and somewhat ad hoc path
selection algorithms means that these systems sometimes lack
sufficient flexibility and path diversity to handle unexpected
situations such as estimation errors, traffic bursts, link failures,
etc.

\jnfpar{Our approach: Semi-oblivious traffic engineering}
This paper investigates a simple question: can we improve traffic
engineering systems by selecting paths in a better way?  We answer
this question positively, showing that by combining \emph{oblivious
routing} (originally formulated by R\"{a}cke~\cite{racke02,racke08})
with dynamic rate adaptation, it is possible to obtain a traffic
engineering system that provides substantial performance improvements
while remaining simple to implement and easy to manage. Moreover,
these benefits continue to hold under a wide range of topologies,
demands, resource budgets, and failure scenarios.  We call this hybrid
approach \emph{semi-oblivious traffic engineering} (SOTE).

\jnfpar{Prior work}
The idea of using oblivious routing for wide-area network traffic
engineering was originally proposed in a seminal paper by Applegate
and Cohen~\cite{applegate03}. They showed that on practical workloads,
oblivious routing performs much better than is predicted by the
worst-case bounds, but they did not investigate whether performance
could be further improved by incorporating dynamic rate adaptation.
Similarly, while the combination of oblivious routing and dynamic rate
adaptation has been explored in the theory literature, prior work
did not develop an implementation and
focused on establishing lower bounds with artificial topologies and
demands~\cite{hajiaghayi07}. This work is the first we are
aware of to describe an implementation of semi-oblivious routing and a
comprehensive evaluation on realistic workloads.

\jnfpar{Implementation}
We have implemented SOTE in a new framework called Kulfi that provides
a rich collection of library functions designed to support rapid
development of traffic engineering algorithms, as well as simulation
and hardware deployments. To date, we have implemented 13 different
traffic engineering algorithms, 7 traffic prediction algorithms, and
evaluated them on 9 different topologies from widely-known ISPs. The
Kulfi simulator includes tunable parameters for demands, failure
scenarios, path budgets, etc. to simulate a variety of realistic
workloads.

\jnfpar{Experience}
The results of our experiments and simulations are extremely
promising. When oblivious routing is enhanced with a rate adaptation
component the congestion ratios we measure are competitive with the
best known traffic engineering solutions, and far better than the
worst-case scenarios predicted in the theory literature. We attribute
these results to qualities of the paths computed by oblivious routing:
they are low-stretch, diverse, and naturally balance load,
guaranteeing worst-case congestion that is within a logarithmic factor
of optimal.

\jnfpar{Contributions}
Our main contributions are as follows:
\begin{enumerate}[nolistsep,topsep=0em,leftmargin=*]
 \item We survey various approaches to traffic engineering in wide
 area networks (\S\ref{sec:background}).
\item We present SOTE, a new approach to traffic engineering that enriches 
  oblivious routing with dynamic rate adaptation (\S\ref{sec:obliv}).
\item We describe an implementation of SOTE and other
  traffic engineering schemes in Kulfi (\S\ref{sec:approach}).
\item We conduct simulations and experiments comparing SOTE
  against other approaches and compare performance with respect to a
  large number of criteria including congestion, latency, throughput,
  loss, resource usage, etc. (\S\ref{sec:hardware}-\ref{sec:evaluation}).
\end{enumerate}

\jnfpar{Outline}
The rest of this paper is organized as follows. The next section
provides a short summary of various approaches to traffic engineering
in wide area networks (\S\ref{sec:background}). We then present a
detailed description of oblivious and semi-oblivious routing
(\S\ref{sec:obliv}). We describe our implementation
(\S\ref{sec:approach}) and present the results of experiments
demonstrating that SOTE is competitive with other approaches on
realistic workloads using an SDN-based hardware testbed
(\S\ref{sec:hardware}) and comprehensive simulations
(\S\ref{sec:evaluation}). Finally, we discuss related work
(\S\ref{sec:related}) and conclude (\S\ref{sec:conclusion}).




\section{Background}
\label{sec:background}

We briefly survey some of the main approaches to traffic engineering,
both in theory and in practice, to lay the groundwork for
understanding SOTE.

\jnfpar{In practice}
The traditional approach to traffic engineering, which has been used
for many years, is to carefully tune link weights in distributed
routing protocols, such as OSPF, so they compute a near-optimal set of
forwarding paths~\cite{fortz-thorup00,fortz-rexford-thorup02}. This
approach is simple to implement, as it harnesses the capabilities of
widely-deployed traditional protocols, but it requires having accurate
estimates of traffic demands. Moreover, it often leads to poor
performance when failures occur or during periods of re-convergence
after link weights are modified to reflect new demands.

Another common strategy is to use equal-cost multi-path routing
(ECMP). Each switch computes a hash over packet headers, and routes
along a randomly selected least-cost path to the destination. Because
forwarding decisions are made without global knowledge, ECMP is simple
to implement, but these local decisions sometimes lead to extra
congestion~\cite{pathak11}. In addition, the performance of ECMP is
fundamentally restricted by its use of least-cost paths, and often
performs poorly in the presence of ``elephant''
flows~\cite{alfares10,alizadeh14}.

Several recent systems have exploited the global visibility offered by
SDN controllers to distribute traffic across pre-computed paths in
near-optimal ways. SWAN~\cite{hong13} distributes flow across
$k$-shortest paths, using a variant of the standard LP formulation
that reserves a small amount of ``scratch capacity'' for configuration
updates. The system proposed by Suchara et al.~\cite{suchara11}
incorporates failures into the LP formulation and computes a diverse
set of paths offline. It then uses a simple local strategy to
dynamically adapt sending rates at each source. B4~\cite{jain13}
distributes flow across multiple paths to improve utilization while
ensuring fairness, and uses a heuristic approximation to improve
scalability.


\jnfpar{In theory}
There is also a significant body of traffic engineering work in the
theory community. One line of work has focused on improving running
times for MCF by relaxing the optimality requirement, and instead
using approximation algorithms, such as the \emph{multiplicative
weights} method~\cite{AHK12,garg2007faster,plotkin95}. For example,
Awerbuch and Khandekar~\cite{awerbuch09} adapted the multiplicative
weights method for distributed settings, improving the scalability of
the approach. However, like all MCF-based schemes, this approach
suffers from sensitivity to estimation accuracy, difficulties related
to fault tolerance, excessive path churn, and high management
complexity~\cite{applegate03}.

To overcome these challenges, another line of work has explored the
space of algorithms that provide strong guarantees in the presence of
arbitrary demands. For example, Valiant load balancing (VLB) routes
traffic in a complete mesh via randomly selected intermediate
nodes. Originally proposed as a way to load balance message routing on
parallel computers~\cite{valiant82}, VLB has recently been applied in
a number of other settings including datacenter networks, wide-area
networks, and software
switches~\cite{argyraki08,dobrescu09,greenberg09,shen08}. However, on
the negative side, sending traffic through intermediate nodes
increases path length, which can increase latency dramatically---e.g.,
consider routing traffic from New York to Seattle through
Europe. Furthermore, VLB can exhibit degraded performance compared to
the optimum when traffic is dropped due to congestion~\cite{dally03}.

Oblivious routing generalizes VLB from meshes to arbitrary
topologies. It computes a probability distribution on low-stretch
paths in advance and forwards traffic according to that distribution
no matter what demands occur when deployed---in other words, it
is \emph{oblivious} to the actual demands. Remarkably, there exist
oblivious routing schemes whose congestion ratio is never worse than
$O(\log{n})$ factor of optimal. The simplest scheme, first proposed in
a breakthrough paper by R\"{a}cke~\cite{racke08}, constructs a set of
tree-structured overlays and then uses these overlays to construct
random forwarding paths between all source-destination pairs.

While the $O(\log{n})$ congestion ratio for oblivious routing is
surprisingly strong for a worst-case guarantee, it still requires that
network operators over-provision capacity by a significant amount.
Applegate and Cohen~\cite{applegate03} showed that, in practice,
oblivious routing performs better than the worst-case predictions, but
a straightforward adoption of oblivious routing is still not
competitive with systems that rebalance load dynamically.
Consequently, the overall verdict seems to be that oblivious routing
is an elegant mathematical result with important applications in the
theory of approximation algorithms (e.g.\ for minimum
bisection~\cite{racke08}), but as a traffic engineering method it is of
limited practical value.



\iftechreport

\newcommand{\cell}[2]{\parbox[c]{#1\textwidth}{\smallskip #2 \smallskip}}
\newcommand{\ncell}[1]{\cell{0.21}{{{\it #1}}}}
\newcommand{\wcell}[1]{\cell{0.3}{{#1}}}
\renewcommand{\arraystretch}{1.5}
\begin{table*}[!ht]
\setlength{\tabcolsep}{3pt}
\begin{center}
\resizebox{\textwidth}{!}{%
    \begin{tabular}{ccccccc}\toprule
        \textbf{Routing Algorithm} & \textbf{Description} & \textbf{Type} & \multicolumn{1}{c}{\centering \textbf{Path}} & \multicolumn{1}{c}{\centering \textbf{Max}} & \multicolumn{2}{c}{\textbf{Overheads}}  \vspace{-6pt} \\
                                   &                      &               & \multicolumn{1}{c}{\textbf{Diversity}}      & \multicolumn{1}{c}{\textbf{Congestion}}      & \multicolumn{1}{c}{\textbf{Churn}} & \textbf{Recovery}  \\ \midrule

\ncell{MCF} & \wcell{Multi-Commodity Flow \\solved with LP~\cite{gurobi}}                           &  conscious        & medium    & least     &  high     &   slow \\ 
\ncell{MW} & \wcell{Multi-Commodity Flow solved\\with Multiplicative Weights~\cite{garg2007faster}} &  conscious        & medium    & least     &  high     &   slow \\ \hdashline
\ncell{SPF} & \wcell{Shortest Path First}                                                           &   oblivious       & least     & high      & none  &   none  \\ 
\ncell{ECMP} & \wcell{Equal-Cost, Multi-Path}                                                       &   oblivious       & low       & high      & none  &   fast  \\ 
\ncell{KSP} & \wcell{K-Shortest Paths}                                                              &   oblivious       & medium    & medium    & none  &   fast  \\ 
\ncell{R\"{a}cke} & \wcell{R\"{a}cke~\cite{racke08}}                                                &  oblivious        & high      & low       & none  &   fast  \\
\ncell{VLB} & \wcell{Valiant Load Balancing~\cite{valiant82}}                                       &   oblivious       & high      & medium       & none  & fast  \\ \hdashline
\ncell{SemiMCF-MCF} & \wcell{MCF for paths\\ MCF for weights}                &   semi-oblivious  & medium    & least    & none & fast \\ 
\ncell{SemiMCF-ECMP} & \wcell{ECMP for paths\\ MCF for weights}              &   semi-oblivious  & low       & medium   & none & fast  \\ 
\ncell{SemiMCF-KSP} & \wcell{KSP for paths\\ MCF for weights}                &   semi-oblivious  & medium    & medium   & none & fast  \\
\rowcolor{gray!20}
\ncell{SemiMCF-R\"{a}cke} & \wcell{R\"{a}cke for paths\\ MCF for weights}    &   semi-oblivious  & high      & low      & none & fast  \\  
\ncell{SemiMCF-VLB} & \wcell{VLB for paths\\ MCF for weights}                &   semi-oblivious  & high      & medium      & none & fast \\ 
\ncell{SemiMCF-MCF-Env} & \wcell{MCF over demand envelope for paths\\MCF for weights~\cite{suchara11}}                  &   semi-oblivious  & medium    & low    &  none & fast   \\
\ncell{SemiMCF-MCF-FT-Env} & \wcell{Multiple MCF-Env considering failures\\MCF for weights~\cite{suchara11}}            &   semi-oblivious  & high      & medium    &  none & fast    \\ \bottomrule
\end{tabular}}
\end{center}
\caption{Summary of routing algorithms implemented in Kulfi.}
\label{tab:algorithms}
\end{table*}%

\else
\fi

\jnfpar{Discussion}
Prior work demonstrates that (i)~traffic engineering algorithms that
use a static set of pre-computed paths can avoid churn and reduce
management overhead, but (ii)~the straightforward adaptation of
oblivious routing requires unacceptable over-provisioning, and
(iii)~although optimizing over the full set of multi-commodity flows is
too heavyweight as a complete strategy, optimizing the distribution of
flow over a limited set of paths (e.g.~using an LP solver, or
iterative methods) appears to work well in practice.  As mentioned
above, several systems have proposed instances of this approach,
including SWAN~\cite{hong13}, B4~\cite{jain13}, and the work of
Suchara et al.~\cite{suchara11}.

\jnfpar{Outlook}
In this paper, we advocate for using oblivious routing as a method for
selecting routing paths, while combining it with optimization
techniques (e.g., LP solvers) to continually adjust the distribution
of flow across those paths. The resulting traffic engineering scheme
has both strong theoretical guarantees and outperforms the state of
the art in simulations of realistic scenarios.



\section{Semi-oblivious Routing}
\label{sec:obliv}


We now present the key building blocks used in SOTE. We first review
R\"{a}cke's oblivious algorithm and then discuss methods for
dynamically adjusting sending rates.

\jnfpar{R\"{a}cke's Algorithm}
At the core of R\"{a}cke's algorithm is a structure we call a {\em
routing tree}. A routing tree determines a unique forwarding path
between every source-destination pair in the network: one simply
concatenates the sub-paths corresponding to edges on the tree path
connecting the pair. In more detail, a routing tree comprises (i)~a
logical tree topology $T$ whose leaves are in one-to-one
correspondence with nodes of the physical topology, $G$; and (ii)~a
mapping that assigns to each edge of $T$ a corresponding path in $G$.
Such a structure implicitly defines a routing path for every
source-destination pair $(s,t)$. One can obtain a path from $s$ to $t$
by finding the corresponding leaves of $T$, identifying the edge set
of the path in $T$ that joins these two leaves, and concatenating the
corresponding physical paths in $G$. Figure~\ref{fig:routing-tree}
illustrates this process of using a routing tree to construct a
physical path.

\usetikzlibrary{fit}

\begin{figure}
\begin{centering}
\iftechreport
\begin{tikzpicture}
[
place/.style={circle,draw=black!50,fill=WildStrawberry!30,thick},
transition/.style={circle,draw=black!50,fill=black!10,thick},
treenode/.style={circle,draw=black!50,fill=cyan!10,thick},
treeedge/.style={dotted},
scale=0.75,
transform shape
]
\else
\begin{tikzpicture}
[
place/.style={circle,draw=black!50,fill=WildStrawberry!30,thick},
transition/.style={circle,draw=black!50,fill=black!10,thick},
treenode/.style={circle,draw=black!50,fill=cyan!10,thick},
treeedge/.style={dotted},
scale=0.45,
transform shape
]
\fi
\definecolor{YellowOrange}{RGB}{255,128,43}
\definecolor{WildStrawberry}{RGB}{233,73,146}
\definecolor{PineGreen}{RGB}{34,170,37}
\node at ( \ax,\ay) (a)[place] {A};
\node at ( \bx,\by) (b)[transition] {B}
edge (a);
\node at ( \cx,\cy) (c)[transition] {C}
edge (b);
\node at ( \dx,\dy) (d)[transition] {D}
edge (c)
edge (a);
\node at (\ex,\ey) (e)[transition] {E}
edge (c);
\node at (\fx,\fy) (f)[transition] {F}
edge (e)
edge (d);
\node at (\gx,\gy) (g)[place] {G}
edge (e)
edge (f);
\node at (\hx,\hy) (h)[transition] {H}
edge (g);
\node at (\ix,\iy) (i)[transition] {I}
edge (f)
edge (h);
\node at (\tcx,\tcy) (tc)[treenode] {C}
edge [blue, \treeedge, bend left=10, line width=2](a)
edge [\treeedge,bend right=5] (b)
edge [\treeedge] (c)
edge [\treeedge,bend left=2] (d);
\node at (\tex,\tey) (te)[treenode] {E}
edge [\treeedge,bend right=5] (e)
edge [\treeedge,bend left=5] (f)
edge [PineGreen,\treeedge,bend right=5, line width=2] (tc)
;
\node at (\tix,\tiy) (ti)[treenode] {I}
edge [red,\treeedge, line width=2] (g)
edge [\treeedge,bend left=20] (h)
edge [\treeedge,bend right=10] (i)
edge [YellowOrange, \treeedge,bend right=10, line width=2] (te)
;

\node at (\ix-0.2, \iy-1.3) {\Large Physical Network};
\node at (\tix+1.7, \tiy+1.5) {\Large Decomposition Tree};
\node at (\tix+1.8, \tiy+1) {\Large Logical path colored};
\node at (\bx-0.7, \ay-0.5)(boundary1) {};
\node at (\hx+0.7, \hy)(boundary2) {};

\node at ( \ax+\shiftamount,\ay) (sa)[place] {A};
\node at ( \bx+\shiftamount,\by) (sb)[transition] {B}
edge [blue!80, line width=2] (sa);
\node at ( \cx+\shiftamount,\cy) (sc)[transition] {C}
edge [blue!80, line width=2](sb);
\node at ( \dx+\shiftamount,\dy) (sd)[transition] {D}
edge (sc)
edge (sa);
\node at (\ex+\shiftamount,\ey) (se)[transition] {E}
edge [PineGreen!80,line width=2](sc);
\node at (\fx+\shiftamount,\fy) (sf)[transition] {F}
edge [YellowOrange!80,line width=2](se)
edge (sd);
\node at (\gx+\shiftamount,\gy) (sg)[place] {G}
edge (se)
edge (sf);
\node at (\hx+\shiftamount,\hy) (sh)[transition] {H}
edge [red!80, line width=2](sg);
\node at (\ix+\shiftamount,\iy) (si)[transition] {I}
edge [YellowOrange!80,line width=2](sf)
edge [red!80, line width=2](sh);

\node at (\ix-0.2+\shiftamount, \iy-1.3) {\Large Physical Network};
\node at (\ix-0.2+\shiftamount, \iy-0.8) {\Large Physical path colored};
\node at (\bx-0.5+\shiftamount, \ay-0.5)(sboundary1) {};
\node at (\hx+0.5+\shiftamount, \hy)(sboundary2) {};

\begin{pgfonlayer}{background}
\node [fill=green!10,fit=(a) (b) (c) (d) (e)
(f)(g)(h)(i)] {};
\node [fill=green!10,fit=(sa) (sb) (sc) (sd) (se)
(sf)(sg)(sh)(si)] {};
\end{pgfonlayer}
\begin{pgfonlayer}{foreground}
\node [fill=olive!10,fit=(a) (b) (c) (d) (e)
(f)(g)(h)(i)(tc)(te)(ti)(boundary2)(boundary1)] {};
\end{pgfonlayer}
\end{tikzpicture}
\caption{A logical representation of 
  a routing tree, and its
  physical mapping.}
\label{fig:routing-tree}
\end{centering}
\end{figure}
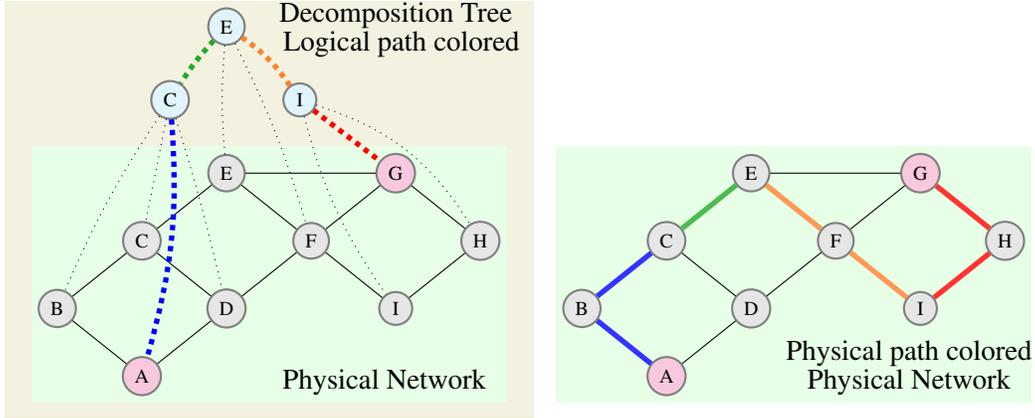

Going a step further, one can define a {\em randomized routing tree}
to be a probability distribution over routing trees. Such a structure
defines a probability distribution over routing paths for
every source $s$ and destination $t$, by randomly sampling one tree
from the distribution and selecting the $s$-$t$ routing path defined
by the sampled tree. As an illustration of the expressiveness of
randomized routing trees, observe that Valiant load balancing (VLB)
can be viewed as a specific instantiation of the general idea. The
logical tree topology in this case consists of a root vertex directly
attached to leaves corresponding to the nodes of the physical
topology. To sample from the probability distribution over routing
trees, one first samples a location for the root vertex at random from
the nodes in the physical topology, and then identifies each tree edge
with the shortest path between the corresponding physical nodes. In
general, one can think of randomized routing trees as a hierarchical
generalization of VLB, where the network is recursively partitioned
into progressively smaller subsets, and routing from a source to a
destination requires passing through a random intermediate node
corresponding to the level of the hierarchy at which the source and
destination are grouped into different pieces of the partition.

R\"{a}cke also proposed an efficient, iterative algorithm for
constructing randomized routing trees. In each iteration, the algorithm
defines a length for each edge of the physical network. Initially the
length of an edge is the inverse of its capacity, and it is updated
multiplicatively at the end of each iteration in a manner to be
described later. For a specified set of edge lengths, the {\em
stretch} of an edge $e=(u,v)$ relative to a routing tree $T$ is
defined to be the ratio between the length of the path from $u$ to $v$
defined by $T$, and the length of the one-hop path defined by $e$. The
{\em average stretch} of a routing tree is computed by averaging the
stretch of each edge, weighted by its capacity. The key subroutine in
R\"{a}cke's algorithm is the FRT algorithm~\cite{FRT03}, which takes
as input a set of edge lengths and outputs a routing tree whose
average stretch is at most $O(\log n)$. Each iteration of R\"{a}cke's
algorithm invokes FRT using the current edge lengths to obtain a
routing tree $T$. Then, for each edge $e$ in the physical topology it
computes a ``utilization'' parameter $u(e,T)$ which constitutes a
worst-case upper bound on the congestion that may be induced on $e$
using $T$ as a routing tree. The routing tree $T$ is then added as a
support point to the distribution, with a probability inversely
proportional to $u_{\max} = \max_{e} \{u(e,T)\}$.  Finally, to update
the length of each edge $e$, we multiply by $(1+\epsilon)^{u(e,T) /
u_{\max}}$, for some constant $\epsilon>0$. In our implementation we
use $\epsilon=0.1$. This loop iterates until there exists an edge $e$
whose combined utilization $\sum_T u(e,T)$ (summing over the trees
selected in all previous iterations of the loop) exceeds a
predetermined threshold.

It is worth pointing out some of the features of the set of paths
selected by R\"{a}cke's oblivious routing algorithm.
First, since each routing tree is a low-stretch tree arising from one
application of the FRT algorithm, R\"{a}cke's oblivious routing is
biased toward low-stretch paths. This contrasts with VLB, which tends
to choose circuitous paths, especially when the source and destination
are near one another.
Second, since FRT is a randomized algorithm, the repeated application
of FRT in R\"{a}cke's algorithm tends to result in a diverse set of
paths for most source-destination pairs. This path diversity is
beneficial for routing around edge or node failures without having to
recompute the entire routing scheme.
Third, in any iteration of the algorithm, the length of an edge is an
exponentially-increasing function of its utilization in previous
iterations.  The inclusion of long edges in routing paths is costly
from the standpoint of constructing a routing tree with low average
stretch. Thus, the routing paths selected in any iteration of
R\"{a}cke's algorithm tend to avoid reusing edges that have been
heavily utilized in prior iterations.  This leads to good
load-balancing properties, e.g.~the $O(\log n)$ congestion ratio
guarantee described in Section~\ref{sec:background}.


\jnfpar{Semi-oblivious routing}
Oblivious routing schemes do not adjust the relative distribution of
flow across paths between a given source and destination. This
limitation means they cannot fine-tune the distribution of flow to
continually re-optimize it as demands evolve. It leads to inefficient
utilization of network capacity and requires overprovisioning to avoid
congestion. Applegate and Cohen~\cite{applegate03} investigated the
performance of oblivious routing in practice and found that, in
contrast the $O(\log n)$ overprovisioning suggested by R\"{a}cke's
worst-case result, in most cases it is sufficient to overprovision the
capacity of each edge by a factor of 2 or less. But while this is
better than the worst-case bounds, it is still not competitive with
state-of-the-art methods.

These observations suggest an investigation of routing schemes that
use a static set of paths (as in oblivious routing) but dynamically
adjust the distribution of flow over those paths as the traffic matrix
varies and/or the network suffers edge or node failures. The
combination of a static set of paths and time-varying probability
distributions over those paths has been called {\em semi-oblivious
routing}~\cite{hajiaghayi07}. Unfortunately, from a worst-case
standpoint, semi-oblivious routing is not significantly better than
fully oblivious routing. Hajiaghayi et al.~\cite{hajiaghayi07} proved
that any semi-oblivious routing scheme that uses polynomially many
forwarding paths must suffer a congestion ratio of
$\widetilde{\Omega}(\log{n})$ in the worst case, even when the network
is as simple as a planar grid. (The actual lower bound
of~\cite{hajiaghayi07} is $\log{(n)}/\log{(\log{(n)})}$, leaving open
the possibility of a very small asymptotic improvement). On the other
hand, the traffic matrices involved in this lower bound construction
are highly unnatural, leaving open the possibility that under
realistic workloads, semi-oblivious routing may significantly
outperform oblivious routing, and may even approach or match the
performance of optimal MCF.

Implementing semi-oblivious routing requires defining two algorithms:
one executed at set-up time to select the static set of forwarding
paths, and another executed repeatedly at run-time to choose how to
distribute flow over those paths.  Table~\ref{tab:algorithms} lists
the algorithms we have implemented in Kulfi including five path
selection algorithms. Three of these (SPF, KSP, ECMP) select shortest
paths or near-shortest-paths, another (VLB) selects paths by
concatenating the shortest paths to a random intermediary node and
then to the destination, and the oblivious scheme (R\"{a}cke) just
described. For distributing flow over paths, in addition to oblivious
schemes (which choose a distribution at initialization time and do not
re-compute at runtime) we have implemented three algorithms for
dynamically adjusting distributions. Semi-MCF uses an LP solver to
minimize the maximum edge congestion when all flow is routed over the
allowed set of paths. Semi-with-MCF-Envelope also uses an LP solver,
but solves a different optimization problem that minimizes a
worst-case upper bound on the congestion under all possible
scenarios~\cite{suchara11}.


\section{Kulfi Toolkit}
\label{sec:approach}

\begin{figure}[t]
\centering
\(\begin{array}{r@{\;}c@{\;}l@{\quad}r@{\;}c@{\;}l@{\quad}r@{\;}c@{\;}l}
T & = &  \textit{Topologies} & P & = & \textit{Paths}  & H & = & \textit{Hosts}  \\
\end{array}\)\\
\(\begin{array}{r@{\;}c@{\;}l@{\quad}l}
\hline
D & \in & (H \times H) \rightarrow  \mathbb{R} & \text{Demand} \\
S & \in & (H \times H) \rightarrow P \rightarrow \mathbb{R} & \text{Scheme}  \\
A  & \in & (T \times D \times S) \rightarrow S  &\text{Algorithm}\\
\hline
\end{array} 
\)
\caption{Domains and Types for Kulfi Modules}
\label{fig:types}
\end{figure}

To facilitate making quantitative comparisons between different
traffic engineering approaches, we developed a software toolkit called
Kulfi. Kulfi provides a set of basic primitives and extensible
mechanisms for rapidly developing and experimenting with a wide
variety of traffic engineering algorithms, both in simulation and
using SDN hardware.

\subsection{Algorithm Modules}

We assume that any traffic engineering algorithm will have access the
network topology (which distinguishes between hosts and forwarding
devices), and a list of demand matrices, where a single such matrix is
a mapping from a source-destination pair to requests for
bandwidth. The demands may be \emph{measured} or \emph{predicted}.

A \emph{routing scheme} is a mapping from source-destination pairs to
distributions on paths between those nodes. For example, a shortest
path routing scheme, given an input $(u,v)$ would return a single path
(i.e., the shortest path from $u$ to $v$) with probability $1.0$. As
we will see, other schemes may return a set of paths, each weighted by
a different probability.

A \emph{routing algorithm} produces a routing scheme. In the simplest
case, the routing algorithm can compute a scheme using only the
topology and demands. However, some routing algorithms must
be \emph{initialized} with an existing scheme to produce a modified
scheme. Thus, in general, a routing algorithm takes a topology, a set
of demands, and a (possibly empty) scheme as input, and produces a
scheme as output. A routing algorithm may be invoked repeatedly, in
response to changes in input (e.g., changes in demands or the
topology).

Kulfi allows users to run different traffic engineering algorithms by
implementing modules that conform to the interfaces in
Table~\ref{fig:types}. These interfaces match the definitions above.
From these definitions, we see there are two axes along which routing
algorithms may differ: how they select paths and how they assign
probabilities to those paths. These two axes lead to three natural
classifications for routing algorithms. An \emph{oblivious} routing
algorithm is one that uses a fixed set of paths and probabilities, and
does not alter those choices in subsequent invocations.
A \emph{semi-oblivious} approach is algorithm uses a fixed set of
paths, but alters the probabilities of those paths with each
invocation. To the best of our knowledge, there is no standardized
term for algorithms that alter the paths and probabilities at each
invocation, so we call these algorithms \emph{conscious}.  The fourth
alternative (i.e., an algorithm that changes paths without changing
probabilities) would not be possible.

We have implemented 13 different algorithms in Kulfi, summarized in
Table~\ref{tab:algorithms}. Among these algorithms, SWAN corresponds
to the Semi-Oblivious with MCF initialized with K-Shortest Paths.
Suchara et al.~\cite{suchara11} corresponds to Semi-Oblivious with MCF
initialized with solving MCF (using an envelope of demands) over all possible failure scenarios. We simulate the inclusion of failures by
repeatedly re-solving on different input topologies. Note that these
two approaches are functionally equivalent for metrics related to
fault tolerance. They differ for metrics related to solving time. But,
as we do we do not include the time spent re-solving as part of the
running time in our simulation, the measurements are slightly
\emph{better} than expected for a direct implementation.

\begin{figure}[t]
\iftechreport
\centerline{\includegraphics[width=0.7\columnwidth]{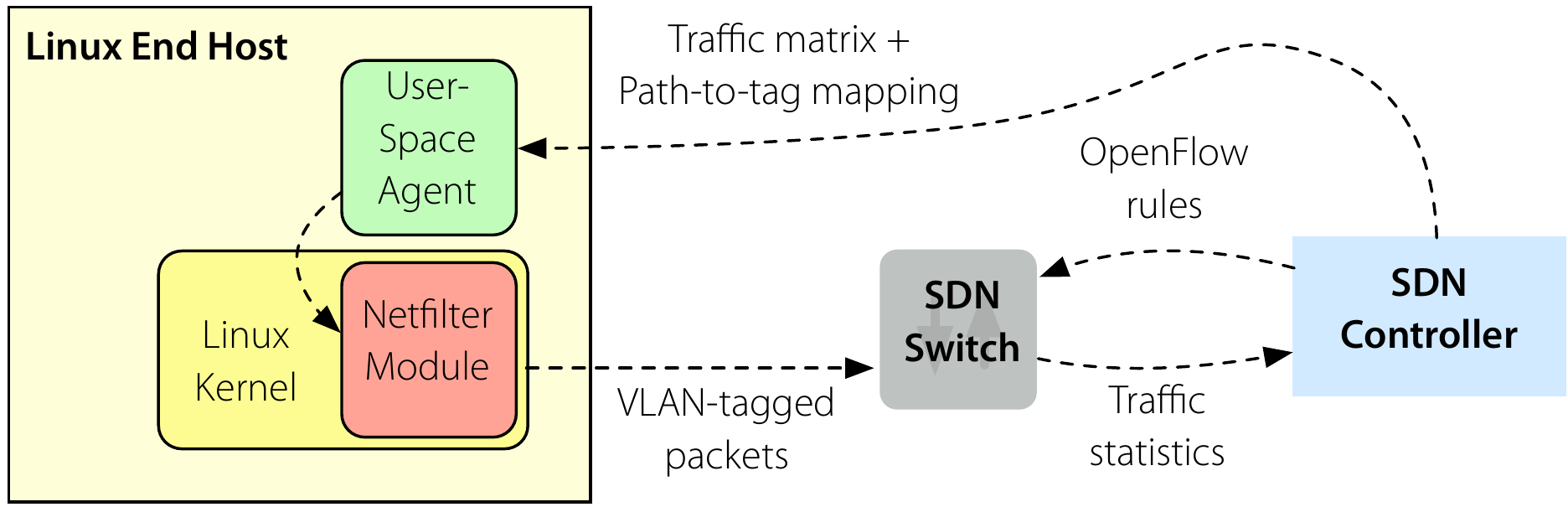}}
\fi
\caption{Implementation architecture.}
\label{fig:architecture}
\end{figure}

\subsection{System Infrastructure}
\label{sec:system}

Kulfi modules can be executed within a simulator, as we will discuss
in Section~\ref{sec:evaluation}, or deployed on actual networks using
a combination of software-defined networking and source-based
routing. This design is inspired by Pathlet routing~\cite{godfrey09}
or Casado's fabric~\cite{casado12}. In our source-routing scheme, each
edge in the network is assigned a unique identifier. A path enumerates
the edges from a source to a destination in a stack of identifiers. To
route, each switch along the path simply ``pops'' the top of the
stack, and forwards out the appropriate port.  It is important to note
that none of the algorithms described in this paper require SDN or
source-routing. However, this approach allowed us to easily implement
and compare many different approaches schemes.



Figure~\ref{fig:architecture} shows the basic architecture for our
implementation. There are four major components, which we will
describe in detail below: (i) an SDN controller, (ii) an
OpenFlow-enabled switch, (iii) an end-host agent in user space, and
(iv) an end-host kernel module.

\jnfpar{SDN controller}
The SDN controller performs four functions. First, it computes the
forwarding paths for the particular traffic engineering scheme (e.g.,
shortest path, optimal MCF, or R\"{a}cke's algorithm). It then maps
the computed paths to their corresponding stack of identifiers in the
physical network. Second, it installs the appropriate forwarding rules
in the SDN switch. Third, it sends the path-to-identifier mappings to
each of the user-space end-host agents in the network.  Finally, it
periodically gathers traffic statistics from the SDN switches,
computes the demands, and sends estimated demands to the user-space
agents on end hosts.

\jnfpar{OpenFlow-enabled switch}
The switches route traffic by examining the identifier stack in the
packet, popping the top of the stack to get the next identifier in the
path, and outputting the packet on the port indicated by the
identifier. Our prototype used VLAN tags to store the stack of
identifiers, although we could have also used MPLS labels. The
switches also maintain counters that collect statistics about the
amount of traffic sent across each link.

\jnfpar{End-host user-space agent}
The end-host user-space agent serves as an intermediary between the
SDN controller and the end-host kernel module. The agent listens on a
designated port for messages from the controller, which contain the
path-to-identifier mappings and a periodically updated global traffic
matrix. The agent communicates this information to the kernel module
through the \texttt{\small /proc} file system.

 \jnfpar{End-host kernel module}
The main responsibility of the kernel module is to assign the
appropriate path to each packet. Packets from the same flow are always
sent along the same paths. Thus, each time the module processes an
outgoing packet, it checks if the packet is part of an existing
flow. If it is not, it assigns the flow a path, and maintains that
information in a hash table. If it is part of a flow, then it
retrieves the path from the table. Flows are evicted from the hash
table based on an idle timeout. For randomized routing schemes, the
kernel module assigns the path using a weighted probability
distribution.






\section{Hardware Experiments}
\label{sec:hardware}

To calibrate the simulations in Section~\ref{sec:evaluation}, we first
conducted experiments on a hardware testbed using the SDN-based
implementation described in Section~\ref{sec:system}. These
experiments measured the congestion ratio for SPF,
ECMP, MCF, R\"{a}cke, and the semi-oblivious variant of R\"{a}cke
using different workloads on an emulated Abilene
backbone~\cite{abilene}.

\jnfpar{Hardware testbed}
Our testbed emulates the Abilene topology (Figure~\ref{fig:abilene}),
which consists of $12$ routers connected by $15$ links, using $3$ Pica8
Pronto 3290 physical switches and $24$ Dell R620 PowerEdge servers. Each
server has two $8$-core $2.6$GHz Intel Xeon processors, $64$GB RAM, and
four $1$Gb NICs. Each switch was configured with $4$ Open vSwitch
instances, yielding a total of $12$ logical switches. We scaled down the
historical Abilene demands to match the $1$Gbps links in our testbed.

\jnfpar{Traffic demands}
We generated traffic using the gravity
model~\cite{roughan2002experience} and also added artificial bursts
over bottleneck links. We used our kernel module to tag outgoing
packets with identifiers corresponding to the path assigned by the
routing scheme. As we replayed traffic in a compressed time-scale, we
configured our kernel module to treat each packet as a separate
flow. These simplifications, which made it easy to conduct careful
measurements, are idealizations, but we believe they are reasonable in
wide-area networks where there are very large number of flows.

\jnfpar{Experimental results}
Figures~\ref{fig:artificial}-\ref{fig:gravity} depict the maximum and
median congestion over all links under different traffic
generators. The first graph (Figure~\ref{fig:artificial}) uses a
traffic pattern in which only switches $4$ (Denver) and $7$ (Kansas
City) sent traffic to switch $10$ (Sunnyvale). This experiment was
crafted so that both flows share a common bottleneck link ($s4$-$s10$)
and the maximum congestion for SPF and ECMP remains $1$ while the
other schemes do not saturate any of the links. Next, we generated
traffic based on
the gravity model of estimating traffic matrices in
Abilene network. Figure~\ref{fig:gravity} shows the
corresponding maximum and median link utilizations. Note that since
the traffic matrices were estimated based on what could be routed
through Abilene network using OSPF, these matrices tend to produce
network traffic that would be feasible under SPF and ECMP. However, we
still observe that the maximum congestion in the case of
semi-oblivious and MCF routing schemes stay well below those for SPF
and ECMP. For the final graph (Figure~\ref{fig:gravityartificial}), we
added the artificial demand on top of the demand estimated by Gravity
model and replayed it on our testbed. Again, this extra demand was not
suited for shortest-paths and the maximum link congestion for SPF and
ECMP shot up while remaining nearly constant for other schemes.

Overall, these experiments show that the semi-oblivious variant of
R\"{a}cke is competitive with optimal MCF approaches in terms of
congestion. Moreover, these results are consistent with the results
that we see in our simulated environment, as we will discuss
Section~\ref{sec:evaluation}.

\iftechreport
\begin{figure}[t!]
\begin{subfigure}[b]{\columnwidth}
  \centering
  \adjincludegraphics[height=0.2\textheight,trim={0 0 0 {0.12\height}},clip]{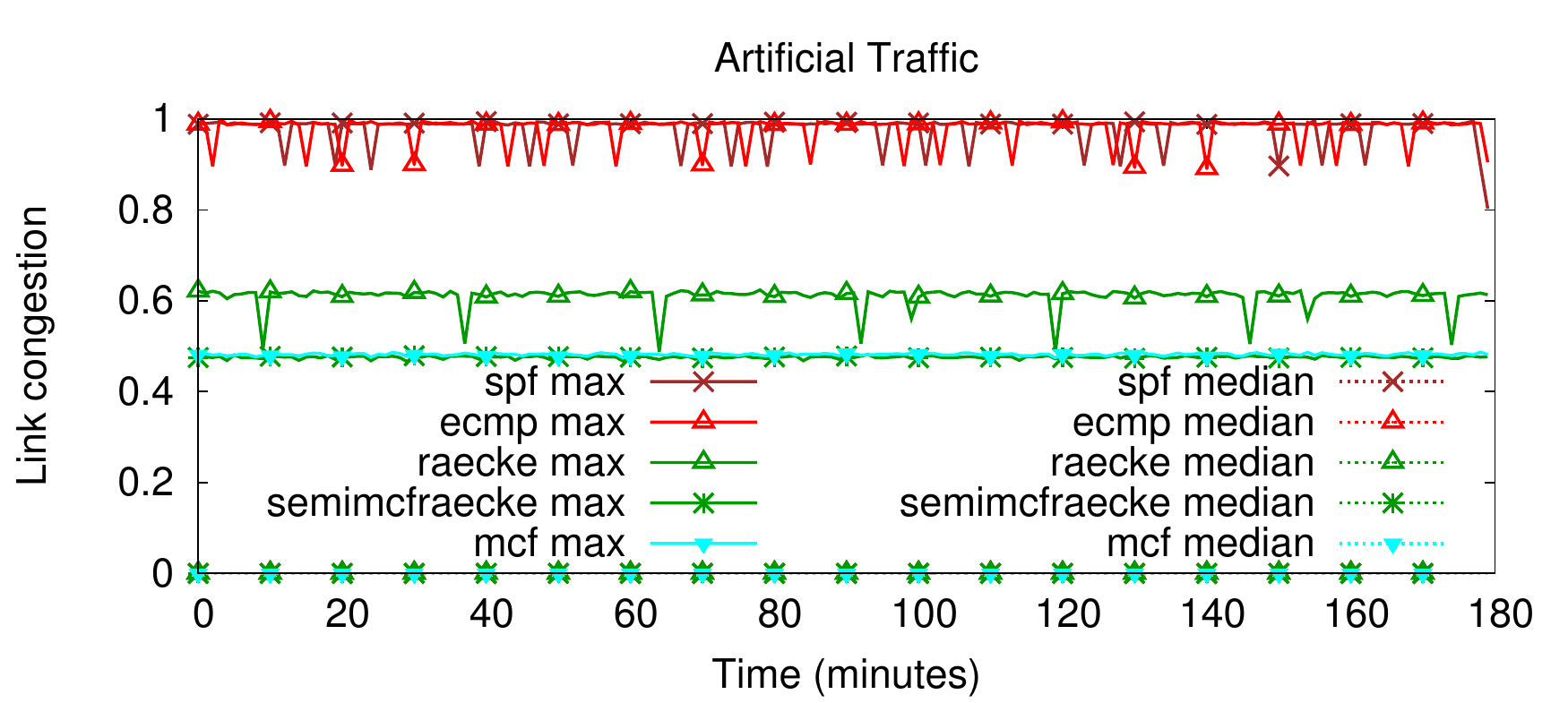}
  \caption{Artificial traffic from switches $4$ and $7$ to $10$.}
  \label{fig:artificial}
\end{subfigure} \par\vfill
\vspace{6pt}
\vspace{6pt}
\begin{subfigure}[b]{\columnwidth}
\centering
  \adjincludegraphics[height=0.2\textheight,trim={0 0 0 {0.12\height}},clip]{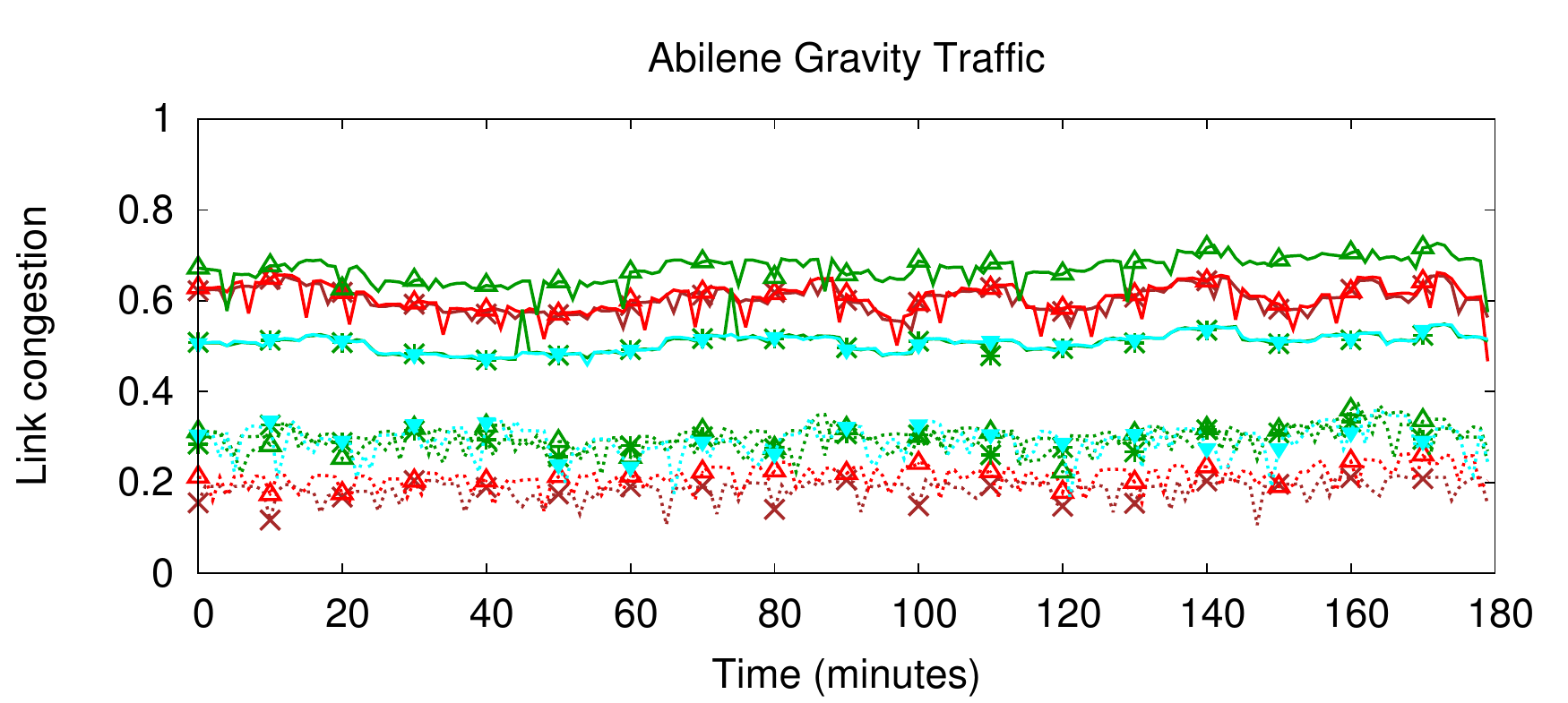}
  \caption{Gravity model demands.}
  \label{fig:gravity}
\end{subfigure} \par\vfill
\vspace{9pt}
\begin{subfigure}[b]{\columnwidth}
  \centering
  \adjincludegraphics[height=0.2\textheight,trim={0 0 0 {0.12\height}},clip]{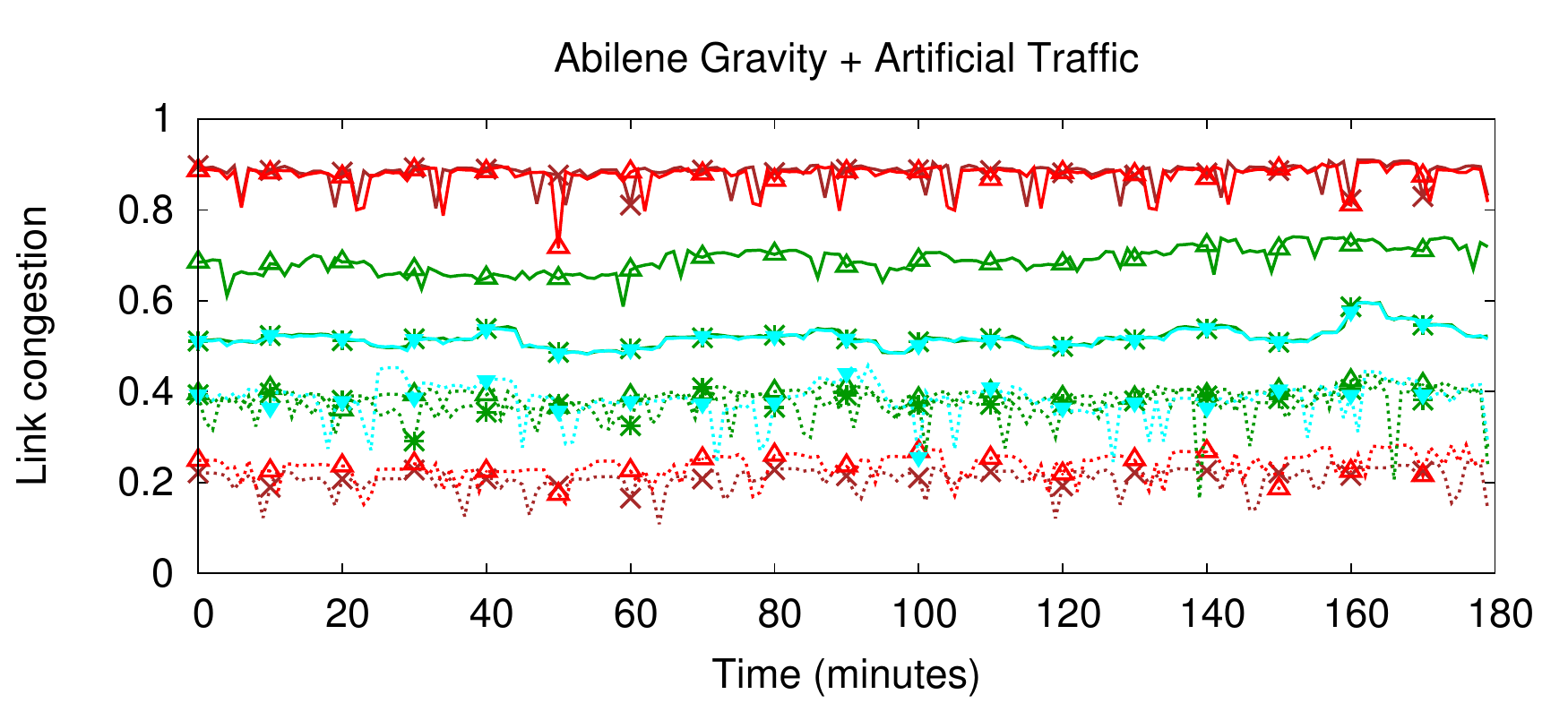}
  \caption{Gravity model demands and Artificial traffic. }
  \label{fig:gravityartificial}
\end{subfigure}
\vspace{5pt}
\caption{Link congestion on emulated Abilene network.}
\label{fig:testbed}

\end{figure}

\fi

\section{Experimental Setup}
\label{sec:setup}

Traffic engineering algorithms are impacted by a number of parameters,
including topology, variance in demands, prediction accuracy, and the
failure model. To provide a comprehensive evaluation, we implemented a
network simulator that allows us to systematically alter the possible
inputs and measure the results.

\jnfpar{Simulator}
The Kulfi simulator is a discrete event simulator that models the
state of a network in response to a sequence of traffic demands. It
has four required input parameters: (i) a file describing the network
topology, (ii) a file containing a list of the \emph{actual} traffic
matrices, (iii) a file containing a list of the \emph{predicted}
traffic matrices, and (iv) a list of algorithms to
execute. Additionally, there are a number of optional parameters
including, among others, flags to vary the budget (i.e., number of
paths used between a source-destination pair), a scaling factor for
demands, the number of links to fail, the recovery method, and the
generation of flash demands.

For each algorithm, the simulator iterates over the sequence of
predicted traffic matrices, which represent the predicted demands for
network sources at discrete time intervals. For each matrix, the
simulator computes the routing scheme and then simulates the flow of
traffic for 1,000 time steps. Every link in the network is associated
with a queue.  At each time step, every source adds traffic to the
queue of the first link of paths it is using. Likewise, every link
forwards traffic to next hop for every flow that it is handling, by
placing traffic on the queue for the next link. Because links have
limited capacity, as specified by the input topology file, the
simulator uses \emph{max-min fair sharing} to allocate bandwidth to
each flow on a link. Any traffic that exceeds the link capacity is
dropped. We refer to traffic dropped due to capacity constraints
as \emph{congestion loss}. During execution, the simulator may fail
some links according to the failure model. If the simulator cannot
forward traffic along a link due to failure, then that traffic is
dropped. We refer to the amount of traffic that is dropped due to link
failure as \emph{failure loss}.


\jnfpar{Topologies}
To ensure that our results are not limited to specific topologies, we
ran simulations on all 262 topologies in the Internet Topology Zoo
dataset~\cite{zoo}. However, to make the presentation of information
accessible, we focus results on a subset of 9
topologies\footnote{Abilene, ATT, British Telecom, GEANT,
Globalcenter, Janet Backbone, NTT, Sprint, Uunet (Verizon)}. We chose the 9
topologies because they are topologies that represent real-world ISPs,
and they overlap with topologies used to evaluate other traffic
engineering approaches in the literature~\cite{kandula05}. To compute
aggregate results, we measured the normalized throughput (and loss due
to congestion or failure) as a fraction of total demand for each
topology and compute averages.


\jnfpar{Demands}
We generated demand matrices using the gravity
model~\cite{roughan2002experience}, which ascribes to each host $i$ a
non-negative weight, $w_i$, and posits that the amount of traffic
flowing from $i$ to $j$ is proportional to the product $w_i \cdot w_j$
for all pairs $i,j$. The weights $w_i$ in our simulations are randomly
sampled from a heavy-tailed Pareto distribution obtained by fitting
the observed Abilene traffic matrices.  To simulate diurnal and weekly
variations in traffic intensity, we rescale the total flow in each
time step based on the weekly traffic patterns observed in NetFlow
traces from Abilene. The patterns are modeled by randomly perturbing
the Fourier coefficients of the observed time-series of total flow
measurements.

To model demand variation over time, we use the Metropolis-Hastings
(M-H) algorithm to sample from a Markov chain on the space of traffic
matrices, whose stationary distribution is the gravity model with
Pareto-distributed weights described above. The M-H algorithm updates
the weight of each host from one time step to the next by randomly
sampling an adjusted value for $w$ using a ``proposal''
distribution. In our simulations we defined the following proposal
distribution for the additive adjustment, $\Delta w$, that
incorporates gradual variation over time accompanied by rare discrete
jumps: with probability 99\%, it is $\mathcal{N}(0,w^2/4)$, and with
probability 1\%, it is uniform$([-w, -0.8w]$ $\cup$ $[0.8w,w])$.

To model flash bursts we induced sudden spikes in demand to a single
sink node in the network followed by a heavy-tailed decrease back to
the stationary distribution. The decreasing tail has a half-life of 30
minutes. In the experiments we pick the sink node randomly in each
traffic matrix (TM). The burst amount is specified by a parameter,
$\beta$. Using $d(\cdot,\cdot)$ to denote entries of the pre-flash TM,
then the peak burst traffic from any host ($h$) to the sink ($s$) is
given by $\beta \cdot \frac{\sum_i \sum_j d(i,j)}{n} \cdot \frac{d(h,s)}{\sum_i d(i,s)} $.

We scale the generated traffic matrices to make them comparable across
different topologies. To do this, we multiply all of the TM's in the
randomly generated sequence by a common scalar, chosen so that the
maximum congestion of an optimal MCF in the first time step is
normalized to 0.4. This choice of the normalization constant is
prompted by recent studies on SWAN~\cite{hong13} which shows that the
average congestion of the maximum utilized links is around 40\% to
60\% in general. However, we also study the routing behavior under
different scales ($S$) for demand. When scaling by $S$, we multiply
all the elements in the traffic matrices such that the expected
congestion for the first traffic matrix is $0.4 \times S$.



\jnfpar{Budget}
In practice, network devices are often constrained in that they can
only support a limited number of forwarding rules. To evaluate how
different routing algorithms are impacted by such resource
constraints, we limited the number of paths that an algorithm can use
between a source-destination pair. We used a budget range from one to
five, as well as unconstrained. For a budget of $k$, we selected the
$k$ paths with highest probabilities and re-normalized their weights.
We refer to the union of the $k$-path sets for all source-destination
pairs as the \emph{base path set}.

\jnfpar{Failure Model}
To evaluate how failures impact various traffic engineering
algorithms, we used several failure scenarios. To create different
scenarios, we varied a parameter $\phi$, ranging from $0$ to $3$,
which specifies the number of failed links. The way in which failed
links were chosen depends on the value of $\phi$. For $\phi>1$, we
simply fail $\phi$ random links as long as it doesn't partition the
network. On the other hand, if $\phi=1$, we choose a deterministic
sequence of single-link failures, to facilitate comparisons across
different routing schemes. The deterministic sequence of failures is
specified as follows:
 we sort the links based on their utilization in SPF and pick links
 from an evenly-spaced sequence of positions in the sorted order. In
 other words, if there are 24 links and 24 TMs, we would end up
 failing each link. If there are 24 links and 12 TMs, we fail
 alternate links in sorted SPF order.

\jnfpar{Failure Recovery}
We implemented two forms of failure recovery. With \emph{global
recovery}, a centralized controller recomputes the routing scheme
based on global knowledge of the network state, and updates forwarding
rules appropriately. With \emph{local recovery}, edge devices (e.g.,
end hosts, virtual switches in the hypervisors, or first-hop switches)
respond to failures without global coordination. To implement local
recovery, we remove the paths that contain the failed links from the
pair's base path set. Then we update the probabilities for the
remaining paths. For SemiMCF-based schemes, these probabilities are
recomputed using MCF, while for other schemes, we renormalize the
probabilities to sum to 1. To implement global recovery, we remove the
failed links from the input topology, and re-compute the scheme. As a
baseline, we also implemented an idealized ``OptimalMCF'' algorithm
that runs MCF every time there is a failure to compute the best scheme
(in terms of congestion) based on real time demands with no prediction
error. For flash crowds, it periodically runs MCF on the current
demands.


\jnfpar{Prediction}
We implemented a suite of algorithms for predicting the next traffic
matrix in a sequence of TMs. These include standard machine learning
methods---linear regression, lasso/ridge regression, logistic
regression, random forest prediction---as well as algebraic methods
(FFT fit and polynomial fit) based on approximating the time series
with a Fourier-sparse or low-degree-polynomial approximation,
respectively. For each pair of hosts, we perform independent time
series prediction. At each time step, every algorithm predicts the
demand of the current time step using the observed demands from the
previous $k$ time steps; the size of the sliding window ($k$) is
optimized separately for each prediction algorithm using
cross-validation. The machine learning algorithms (regression
algorithms and random forest) are trained using the previous $k$ time
steps as $k$ features. The FFT fit and polynomial fit algorithms find
a function with a bounded number of non-zero Fourier coefficients (FFT
fit) or a bounded-degree polynomial function (polynomial fit) that
minimizes the absolute difference between the predicted demand and the
actual demand over the past $k$ time steps. This best-fit function is
then evaluated in the current time step to yield the predicted demand.
Polynomial fit and random forest are unstable, especially when spikes
exist. Regression algorithms with regularizers, such as lasso and
ridge, are more stable. But the prediction errors are usually less
than 20\% of the actual demand, both on real Abilene data and on the
synthetic demand matrices that we generated.

\jnfpar{Prediction Error}
We tested the sensitivity of various traffic engineering approaches to
prediction errors. We first generated a sequence of synthetic demand
matrices representing the actual traffic demands between hosts. We
then generated a simulated sequence of noisy predictions by performing
random perturbations of the actual demands.  More specifically, for
each host and each time step, we perturbed its weight in the gravity
model by multiplying the weight by $1+\epsilon$ or $1-\epsilon$ with
equal probability; then we generated the predicted TMs based on the
perturbed weights. Simulating predictions in this way (rather than by
running the aforementioned prediction algorithms) allows us to
directly modulate the error parameter ($\epsilon$) in order to assess
the sensitivity of different routing algorithms to prediction error.

\section{Evaluation}
\label{sec:evaluation}

To evaluate SOTE, we ran simulations on 13 different traffic
engineering algorithms, 7 different prediction algorithms, and 262
different topologies. We experimented with a variety of traffic
patterns, scale of demands, failure scenarios, error in demand
prediction, and budget constraints. We collected measurements for
throughput, congestion, failure loss, latency, number of paths used,
solver time, and path churn.

This section presents only a small subset of our experimental data
that illustrates a few key results. Unless otherwise specified, the
data reported is an average of results from the 9 ISP topologies, with
the synthesized traffic matrices using the gravity model. The first
two experiments present results using the same Abilene topology as
in \S\ref{sec:hardware}.

\begin{figure*}[t]
\iftechreport
\centerline{\includegraphics[width=\columnwidth]{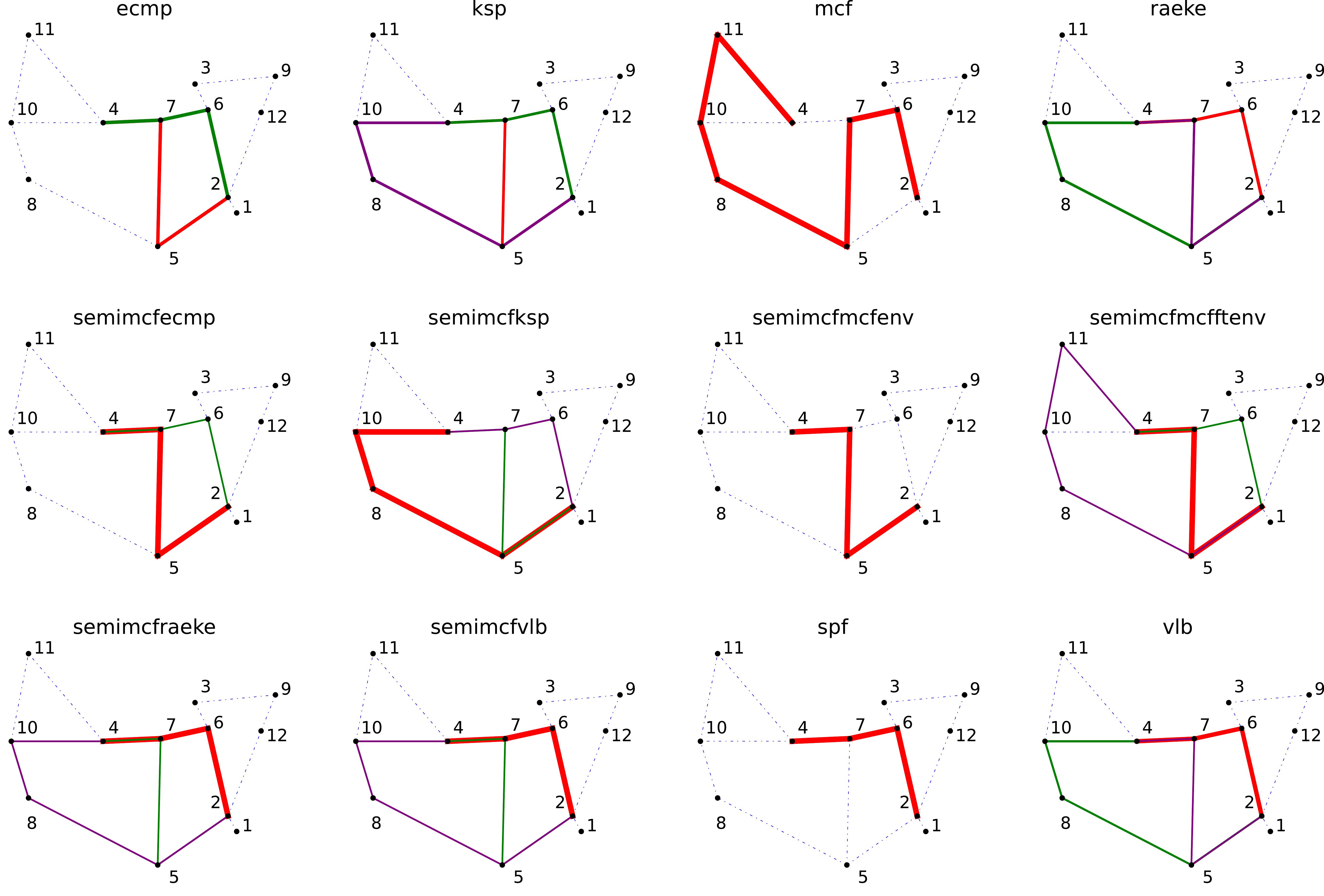}}
\fi
\caption{Case study: Abilene topology showing h4$\rightarrow$h2 paths used by different routing algorithms.}
\label{fig:abilene}
\end{figure*}


\iftechreport
\begin{figure}[t]
\centerline{\includegraphics[width=\columnwidth]{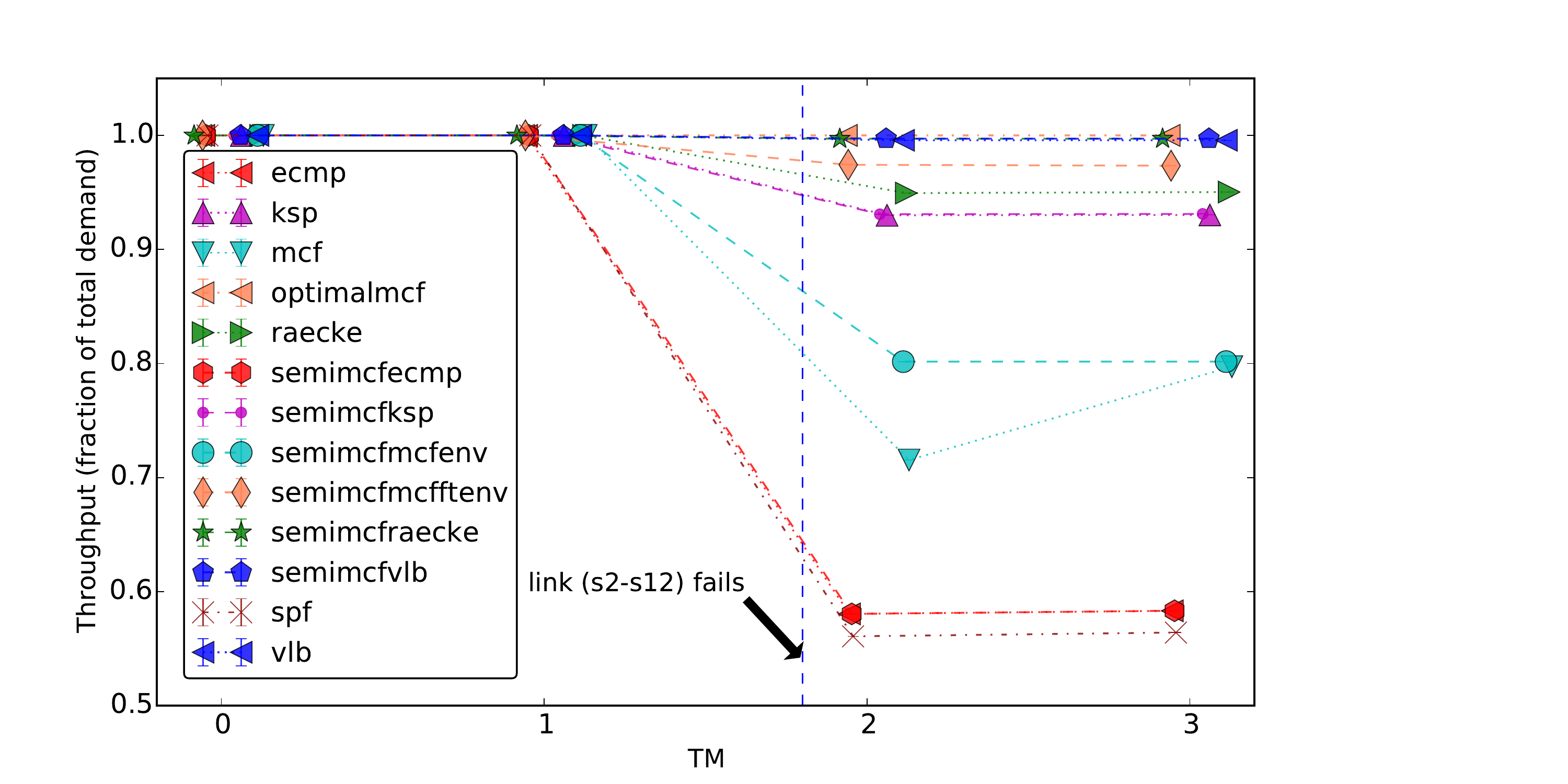}}
\caption{Case study: Throughput as fraction of total demand when link \textit{s2-s12} fails.}
\label{fig:case-tput}
\end{figure}
\fi

\textbf{Case Study: How diverse are the paths chosen by different algorithms?} 
A central thesis of this paper is that path selection algorithms have
a large impact on the performance of traffic engineering schemes. We
first explored exactly how different are the path sets selected by
different algorithms. To visualizes the differences, we depict only
the paths from a single source, PoP 4 (Denver), to a single
destination, PoP 2 (Atlanta). We enforce a path budget of 3. In
Figure~\ref{fig:abilene}, the red (thick) paths have the highest
weights while the purple (thin) paths have the lowest weights. It is
evident that the paths selected by different schemes may vary
significantly on different runs.

While some paths are obvious (e.g., SPF picks the shortest path),
others are sometimes surprising.  For example, the
``congestion-optimal" MCF routing scheme uses a very long path. Since
MCF aims to minimize the maximum congestion, and not reduce latency,
it may select long paths to accommodate other demands in the
network.

Another interesting observation is that the path weights may differ
significantly between oblivious schemes and their semi-oblivious
alternatives, such as in R\"{a}cke and SemiMCF-R\"{a}cke. Moreover,
even two semi-oblivious algorithms initialized with the same set of
paths may differ in path weight. This is because when adapting rates
to minimize congestion, the set of paths and demands between other
pairs of PoPs are also considered. For example, SemiMCF-KSP assigns a
higher weight to a longer path, because if it used the shorter one, a
higher max-congestion would have occurred.

\textbf{Case Study: Does path selection impact throughput when a link fails?} 
To evaluate how path selection impacts performance in the presence of
failures, we ran each algorithms under a simple failure scenario. We
performed routing for four traffic matrices and failed the link
connecting PoPs numbered 2 (Atlanta) and 12 (Washington) for traffic
matrices 2 and 3. Figure~\ref{fig:case-tput} shows the fraction of
total demand that could be delivered successfully. To recover from
failure, each algorithm deployed local recovery. Since SPF selects
only one path for any pair of PoPs, there is no way it can recover
from failure and hence the throughput decreases drastically. On the
other hand, schemes that have a large set of diverse paths, such as
SemiMCF-R\"{a}cke, are more fault tolerant.

\begin{figure}[t]
\centerline{\includegraphics[width=\columnwidth]{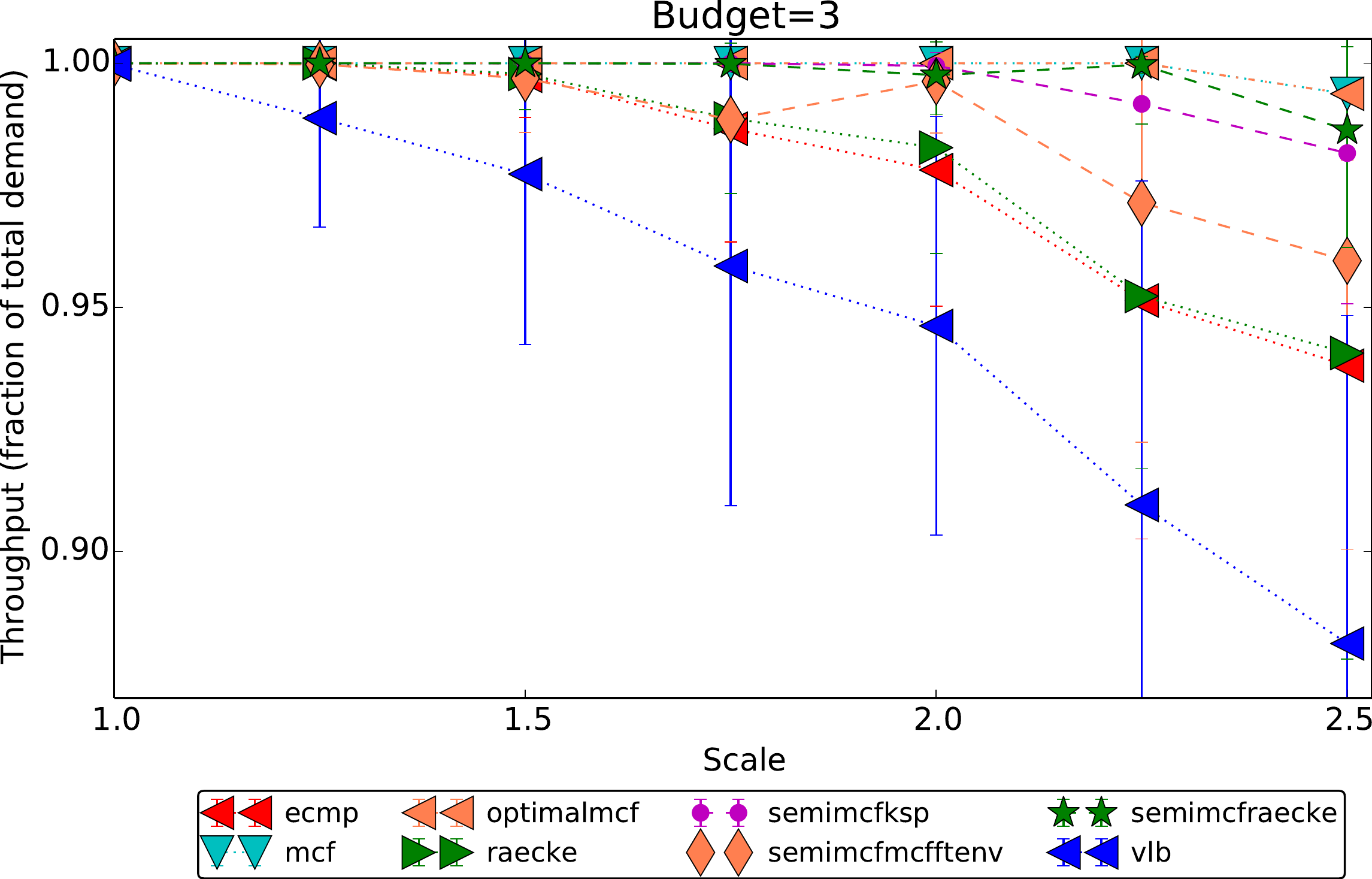}}
\caption{Impact of scaling demands by $S \in [1, 2.5]$: throughput decreases and congestion loss increases.}
\label{fig:demandscale}
\end{figure}

\textbf{How does increased load impact throughput and loss?}
To evaluate how path selection affects performance under a variety of
demands, we scaled the synthetic traffic matrices by a configurable
factor $S$, for increasing values of $S$ from 1 to 2.5. The
SWAN~\cite{hong13} paper reports that 0.4 is the maximum average
congestion in real networks. For this reason, we define $S=1$ as
corresponding to those traffic matrices for which the minimum
max-congestion is 0.4. $S=2$ are matrices for which the minimum
max-congestion is 0.8, etc. We measured the throughput, defined as
fraction of the total demand, and congestion loss for all algorithms
on the set of 9 topologies as described
earlier. Figure~\ref{fig:demandscale} shows the expected trend that
congestion loss increases with increasing scale. VLB, owing to the
longer expected path lengths, starts experiencing maximum congestion
loss as the network reaches capacity.
For $S > 2.5$, congestion loss is unavoidable as the minimum possible max-congestions reaches 1.
We observe that:
\begin{enumerate}
    \item MCF-based schemes perform the best as they are optimized to minimize max-congestion and achieve optimal performance.
    \item \emph{SemiMCF-R\"{a}cke provides throughput close to optimal and better than all other non-MCF based schemes.}
    \item SemiMCFMCFFTEnv also takes failures into account and thus, it can select longer paths which do not minimize loss due to congestion.
\end{enumerate}

%

\textbf{How do link failures impact throughput and loss?}
\iftechreport
Next, we measured the throughput and loss for the various algorithms
as we: (i) increased the number of links ($\phi$) that can fail
simultaneously from 0 to 3, (ii) varied the budget from 3 to 5, and
(iii) increased the scale factor $S$ from 0.5 to 4.  The failures are
picked randomly, provided they don't disconnect the network.
We show the loss due to failure as a fraction of the total demand in
figure~\ref{fig:failures_bar}. The y-axis is failure loss, and
the x-axis is scale factor. Each graph shows the results for a fixed
number of failed links. The budget was set to 3 for the 3 graphs in the top half
, and it was set to 5 for the other 3.

\begin{figure}[]
\clearpage
\begin{subfigure}[b]{\columnwidth}
    \centerline{
    \adjincludegraphics[width=0.91\columnwidth]{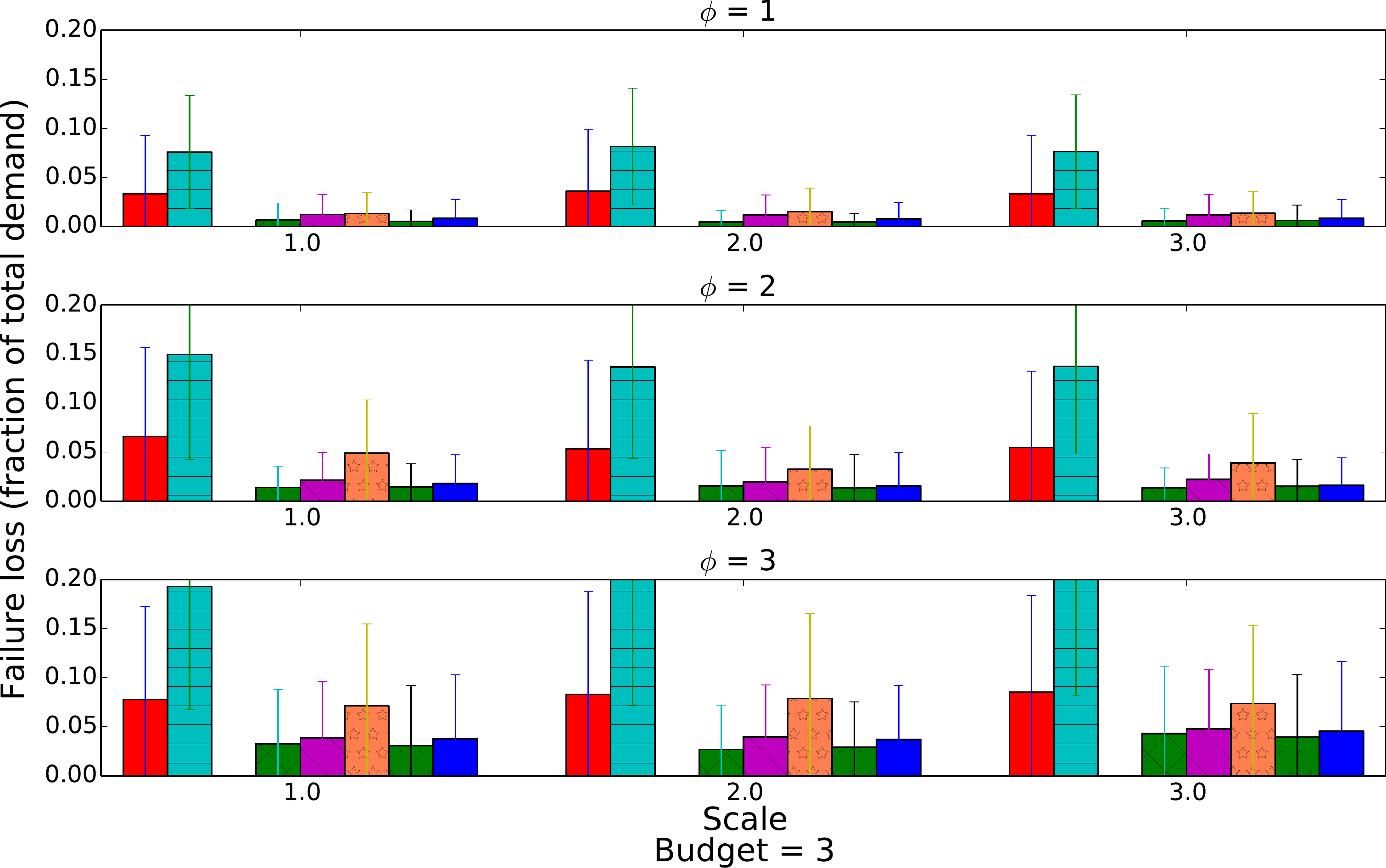}}
\label{fig:failures_b3}
\end{subfigure}
\begin{subfigure}[b]{\columnwidth}
    \centerline{\adjincludegraphics[width=0.92\columnwidth]{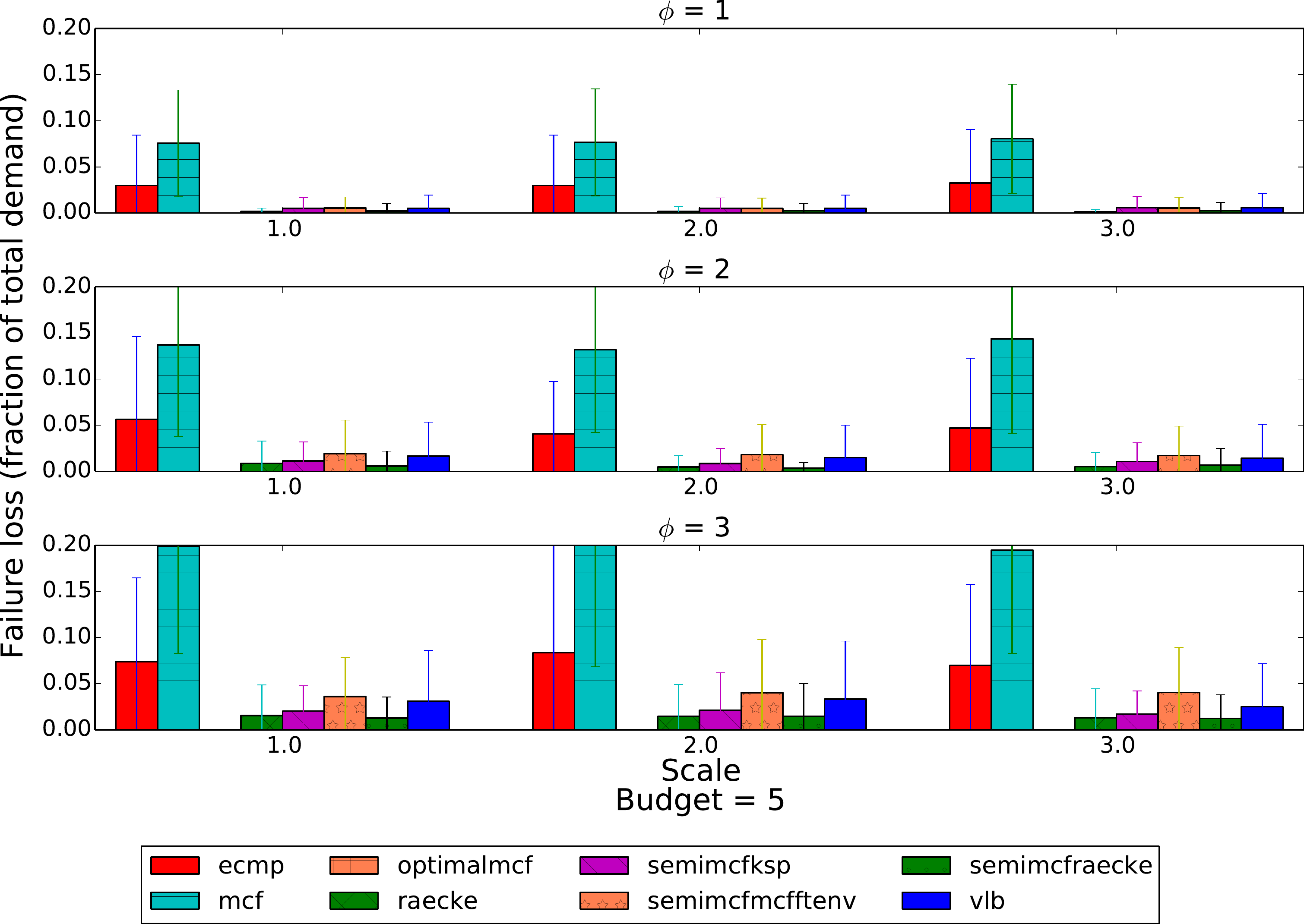}}
\label{fig:failures_b5}
\end{subfigure}
\caption{Failure loss under different failure scenarios, scale, and budget. (a) budget=3 (b) budget=5}
\label{fig:failures_bar}
\clearpage
\end{figure}

We observe the following:
\begin{enumerate}
    \item \emph{In almost all scenarios,
R\"{a}cke and SemiMCF-R\"{a}cke have the minimum loss due to failure.}
    \item Schemes that utilize the full budget, such as KSP, R\"{a}cke
and VLB, experience less traffic loss. 
    \item As the number of
failures increase, schemes that don't use a diverse set of paths, such
as SPF and MCF, react poorly.
\end{enumerate}

\fi


\textbf{How does prediction accuracy impact throughput?}
To evaluate the sensitivity to prediction error, we define a parameter
$\epsilon$ as the factor used to perturb the weights of hosts while
estimating demands based on the gravity model. We vary $\epsilon \in
[0.00, 0.80]$ to generate traffic matrices which are used as predicted
traffic matrices by the routing algorithms.
\iftechreport
Figure~\ref{fig:prediction_error_maxcong} shows the effect on
maximum congestion caused by this error in prediction. Oblivious routing
schemes, as they don't depend on predicted demands, don't seem show
any noticeable change with increasing prediction errors. However,
algorithms which are MCF-based or SemiMCF-based (use predicted
demands) show significant increase in congestion as prediction error increases.
MCF starts with same maximum congestion as OptimalMCF at $\epsilon = 0.0$, but 
quickly deteriorates with $\epsilon$.\\
\emph{Semi-oblivious schemes are more resilient compared to MCF as their base path set do not
change for different $\epsilon$}. Unlike MCF which can select a completely different
set of paths for different $\epsilon$, these schemes perform rate adaptation over the
same base path set as with $\epsilon = 0.0$.\\

\fi

\iftechreport
\begin{figure}[!ht]
\centerline{\includegraphics[width=0.92\columnwidth]{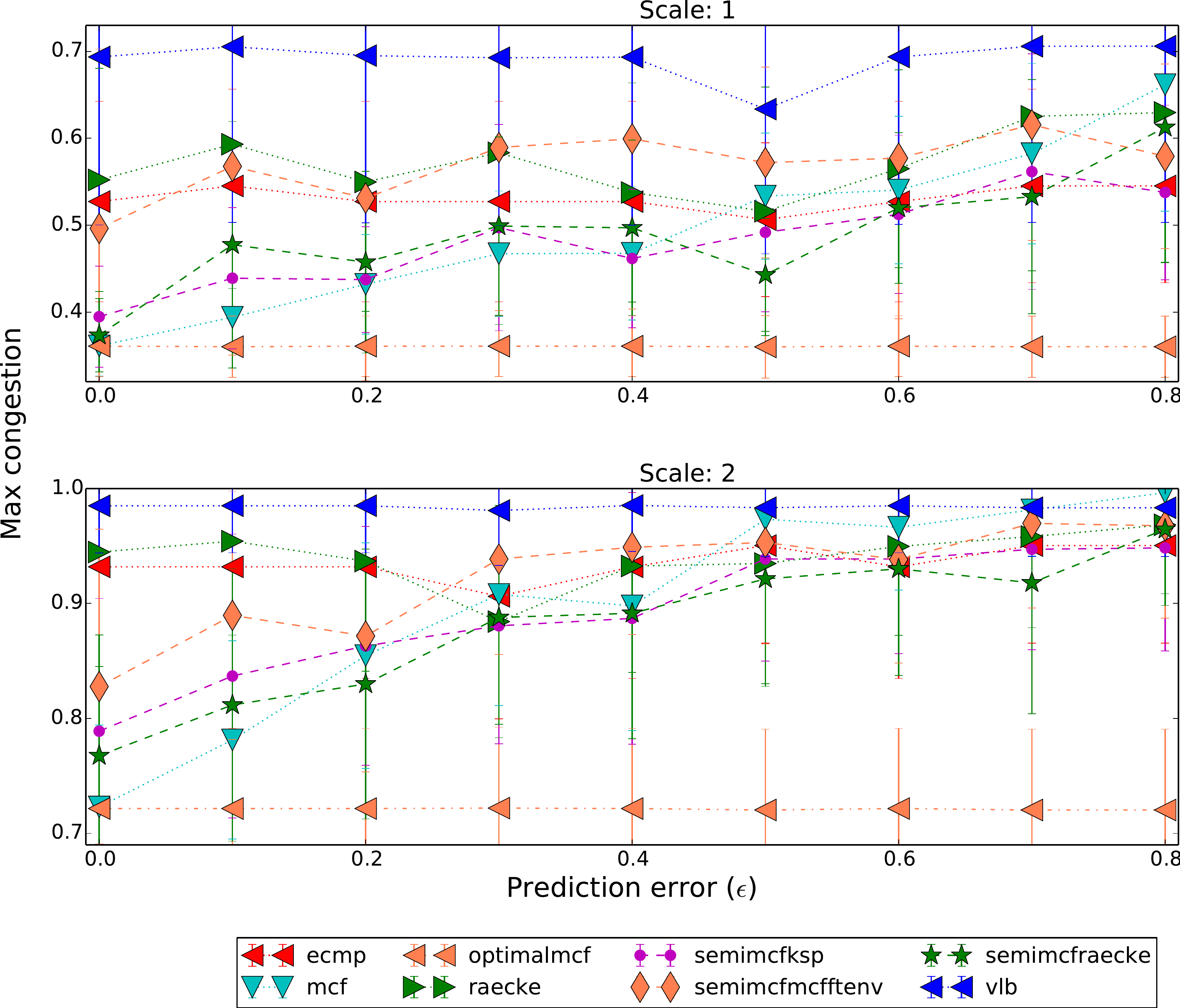}}
\clearpage
\caption{Resilience to error in demand prediction.}
\label{fig:prediction_error_maxcong}
\end{figure}

\fi

\iftechreport
\begin{figure}[]
\clearpage
\centerline{\includegraphics[width=\columnwidth]{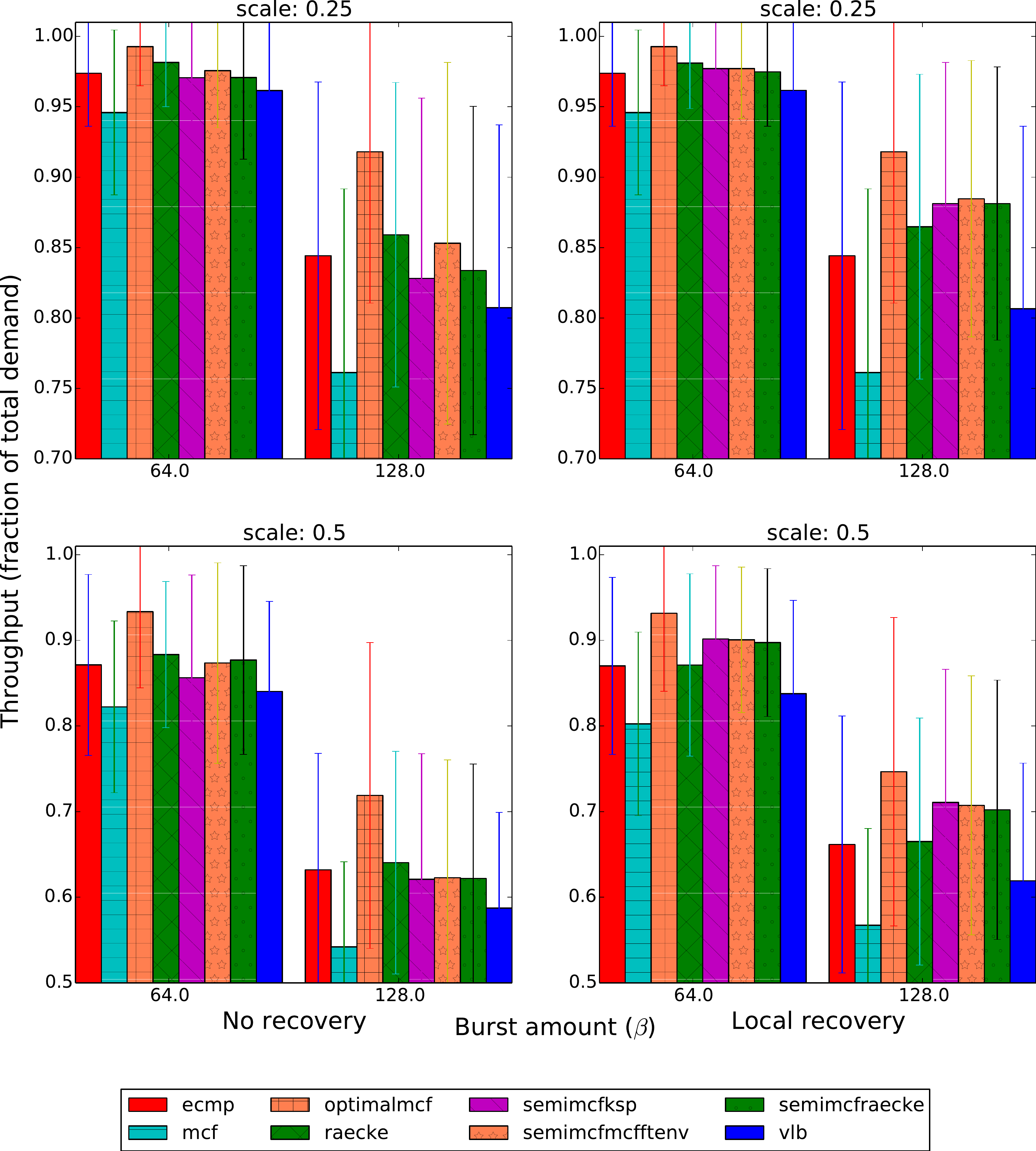}}
\clearpage
\else
\begin{figure}[t]
\centerline{\includegraphics[width=\columnwidth]{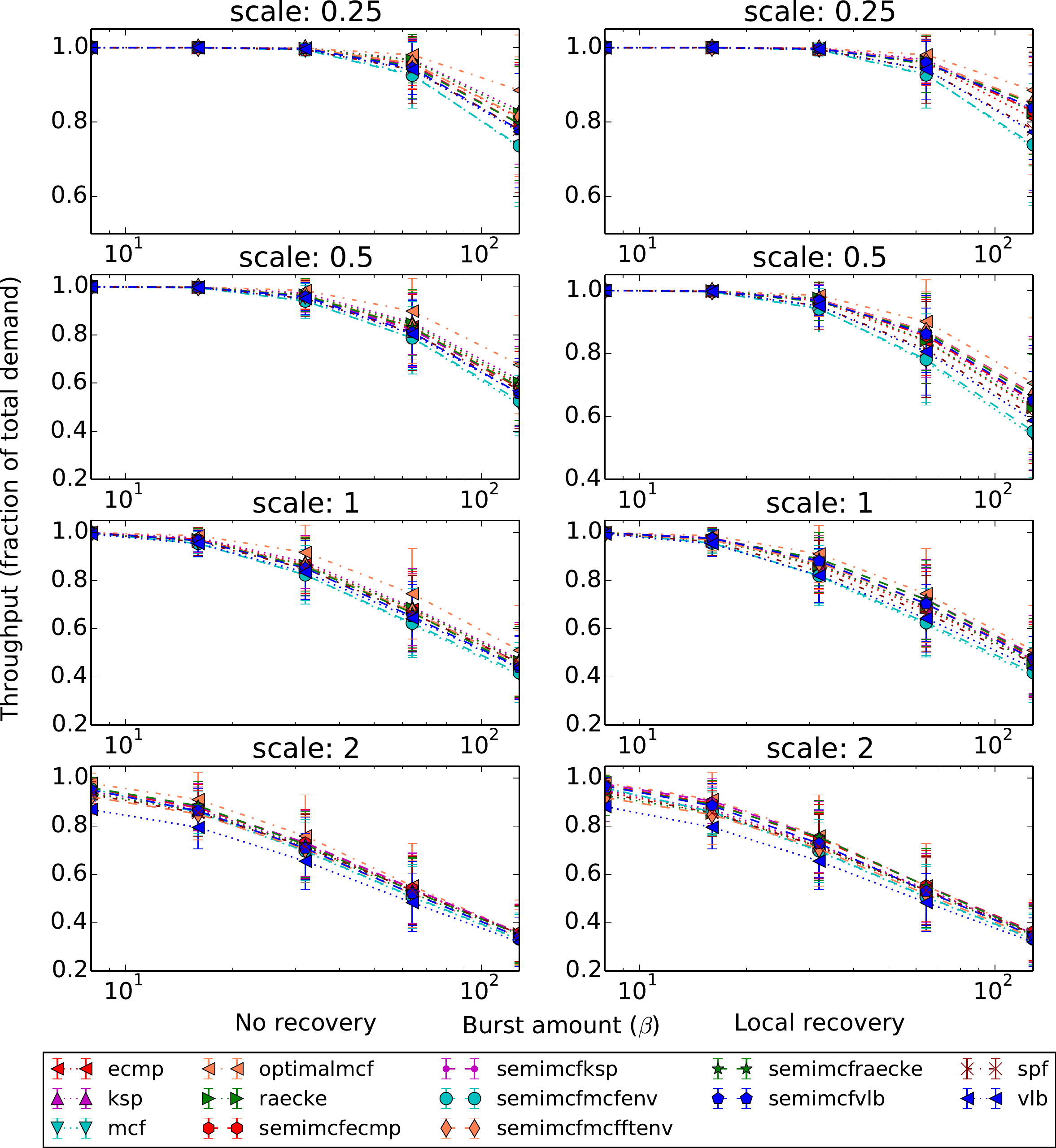}}
\fi
\caption{Effect of flash bursts on throughput. Semi-oblivious schemes improve with local recovery.}
\label{fig:flash}
\vspace{4pt}
\end{figure}

\textbf{How do flash traffic bursts impact throughput?}
We compare the performance of different algorithms under flash bursts
as we vary the amount of burst ($\beta$). We select a random sink
hotspot (same for every algorithm) for every traffic matrix and
measure the effect of congestion. Every algorithm has a flash recovery
mechanism similar to ``local recovery''. It is based on actual traffic
matrix with a lag $\delta$. So, at time $t$, the algorithms have
access to traffic observed at $t-\delta$. In our experiments, we use
$\delta = 8$. As the flash demands decay over time, in our
experiments, all the algorithms perform local flash recovery
(OptimalMCF does global recovery) every 200 time steps.

We measured the throughput as we increased the burst factor, $\beta$.
Figure~\ref{fig:flash} shows the results. The graphs in the left
column show the throughput without the recovery mechanism. The graphs
on the right show the throughput with recovery. OptimalMCF is
omniscient and uses actual traffic demands at the current time to
compute a routing scheme by solving MCF. Thus, it achieves the best
congestion ratio and high throughput. It easily outperforms other
routing algorithms as long as there is spare capacity in the network
(holds for $S \leq 2$). However, on enabling flash recovery, we find
that most of the SemiMCF-based schemes improve in terms of
throughput. MCF and SemiMCFMCFEnv deteriorate more than others as they
are susceptible to inaccuracy in demand prediction. \emph{SemiMCF-R\"{a}cke
and SemiMCFMCFFTEnv perform the best with local recovery}. SemiMCFMCFFTEnv
differs from MCF and SemiMCFMCFEnv because when it chooses a more
diverse set of paths to be more resilient to failures, it
automatically gets more path diversity to avoid bottlenecked links
during flash crowds. The same reason holds for SemiMCF-R\"{a}cke.

\textbf{How does path selection affect latency?}
Figure~\ref{fig:latency} shows the CDF of latency for three of the
largest topologies we ran our experiments on. We see similar results
with smaller topologies.  We select the scale to be 2.5 so that the
minimum max-congestion is 1.0, and we notice the effect of congestion.
\begin{enumerate}
    \item SPF, KSP, ECMP and their SemiMCF-based variants use relatively shorter
paths and thus have better 80-th percentile latency. But, since they
don't distribute traffic efficiently over less congested links, they
incur higher tail latencies.
    \item In contrast, SemiMCFMCFEnv and OptimalMCF
can successfully deliver all the traffic but use longer paths.
    \item \emph{SemiMCFR\"{a}cke occupies an attractive middle ground.
It has similar 80-th percentile latency as the SPF, KSP and ECMP-based
schemes. And it can also deliver the remaining 20\% traffic using
slightly longer paths as it distributes the traffic more efficiently.}
    \item SemiMCFMCFFTEnv trades off latency for more diverse (fault tolerant)
but longer paths.
\end{enumerate}

\iftechreport
\begin{figure}[!h]
\centerline{\includegraphics[height=0.92\textheight]{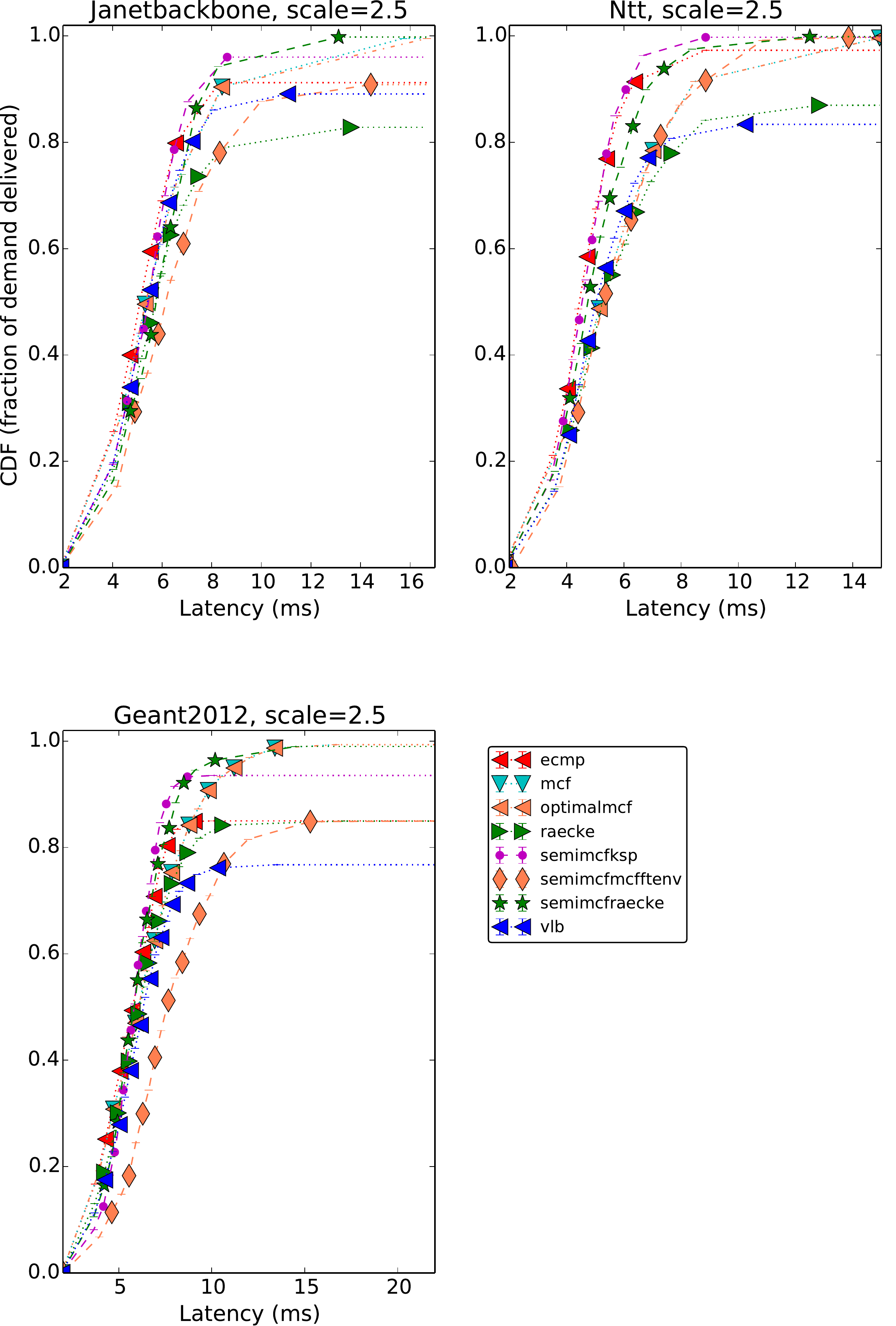}}
\clearpage
\else
\begin{figure}[t]
\centerline{\includegraphics[width=\columnwidth]{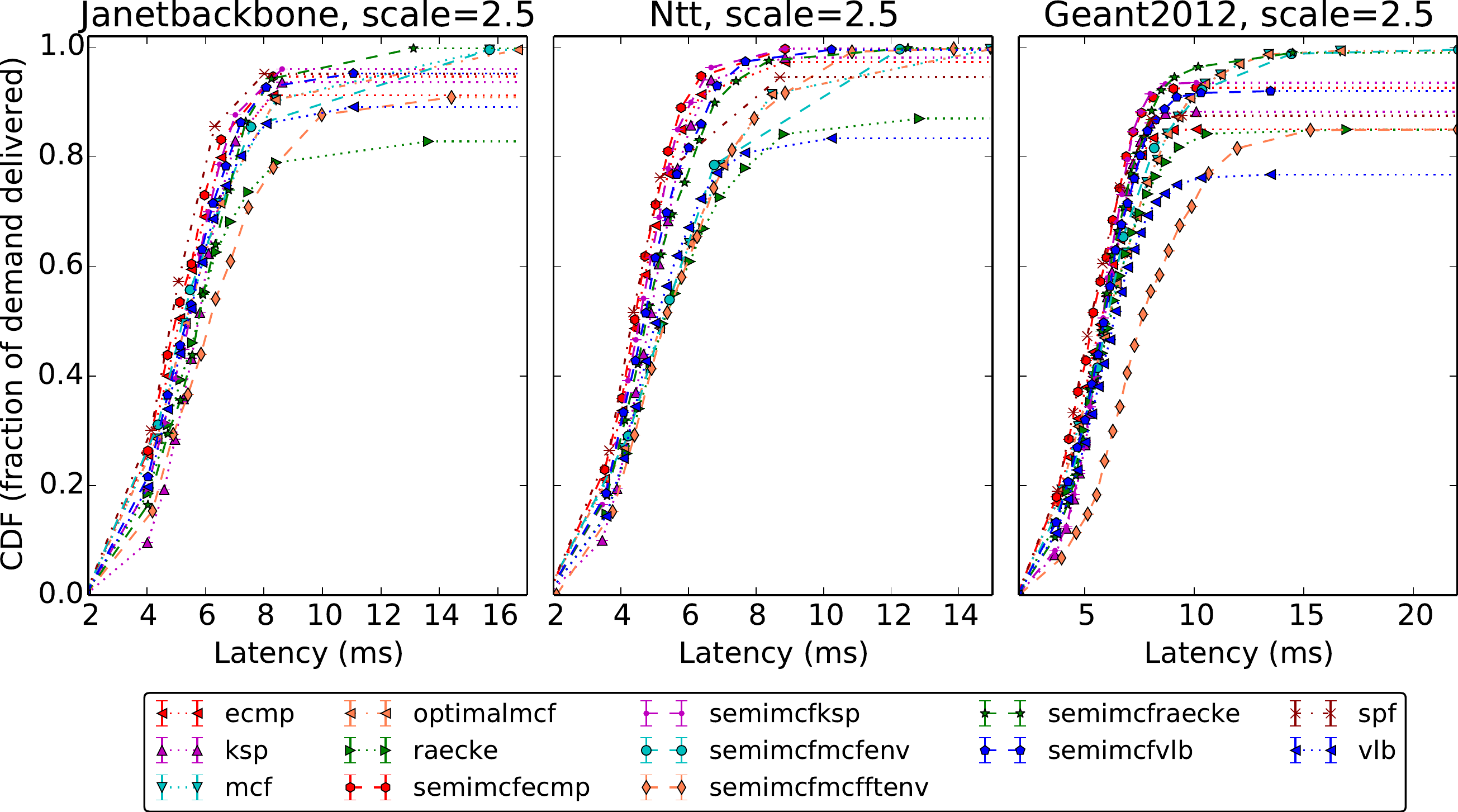}}
\fi
\caption{Latency CDF for different topologies.}
\label{fig:latency}
\end{figure}

\textbf{How does path budget impact throughput and loss?}
SDN switches can only install a limited number of forwarding rules
bound by their TCAM size. Routing schemes that use an excessive amount
of rules may not be deployable in practice. For a network with $n$
vertices and $m$ edges, R\"{a}cke's oblivious routing scheme can use
$O(mn^2 \log n)$ paths, or $O(m\log n)$ paths on an average between
each pair of source and destination. Similarly, VLB and MCF can use up
to $O(n)$ and $O(m)$ paths per pair, respectively. To address the
limitations on number of forwarding rules, we enforce a budget on each
algorithm. The budget restricts the number of paths used for each
source and destination pair. Figure~\ref{fig:budget} shows the effect
of the budget as demand scales. Initially, increasing the budget
boosts throughput for SemiMCF-based schemes initialized with
R\"{a}cke, MCF and VLB, as these schemes are able to spread traffic
over a greater number of paths and able to minimize
congestion. However, the improvement with increasing budget becomes
negligible beyond a budget of 5. \emph{This shows that even though routing
algorithms like R\"{a}cke's need a lot of paths in theory to be
efficient, they need only a reasonably small constant number of paths for
real world WANs.}

\iftechreport
\begin{figure}[]
\clearpage
\centerline{\includegraphics[width=\columnwidth]{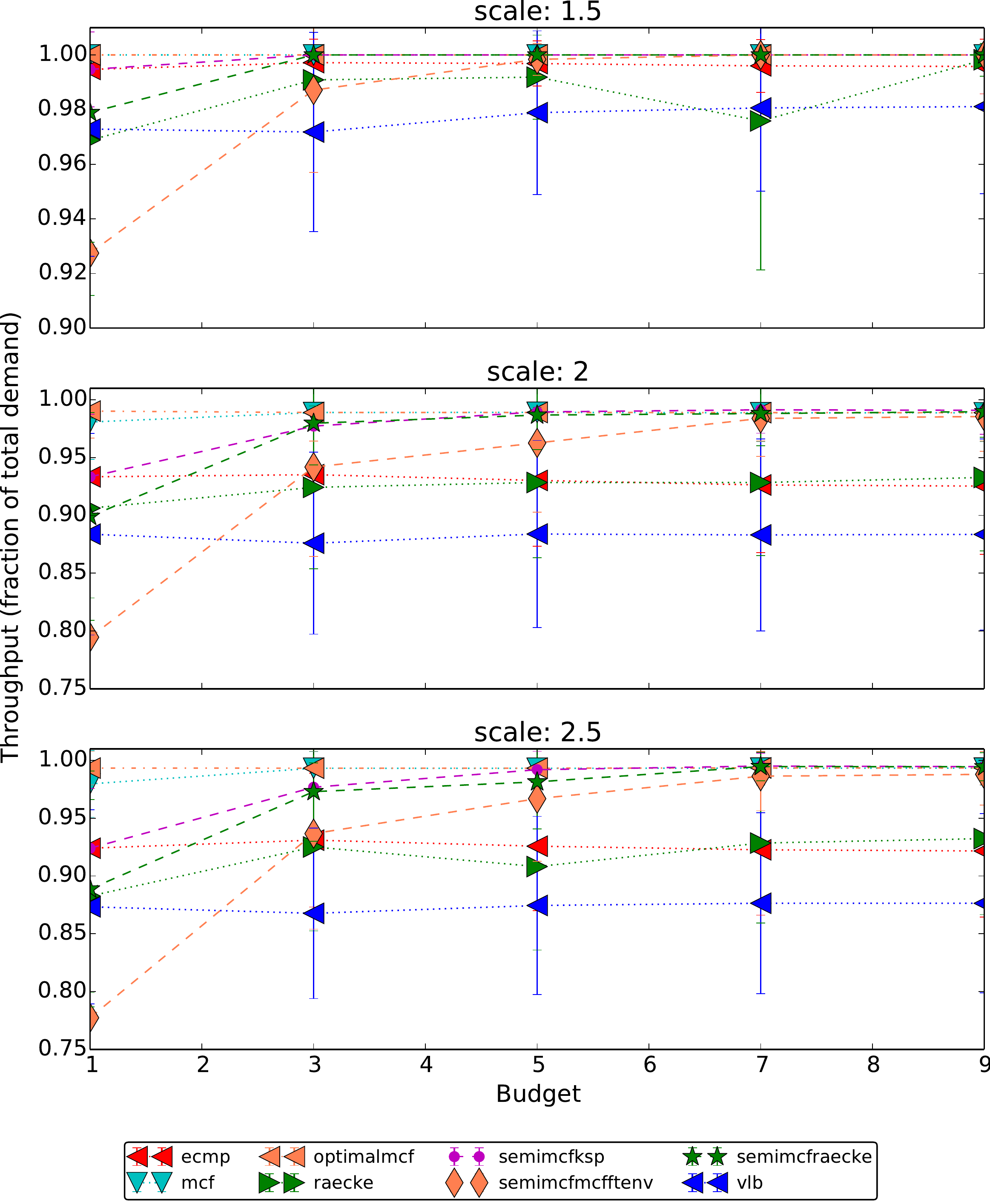}}
\clearpage
\else
\begin{figure}[t]
\centerline{\includegraphics[width=\columnwidth]{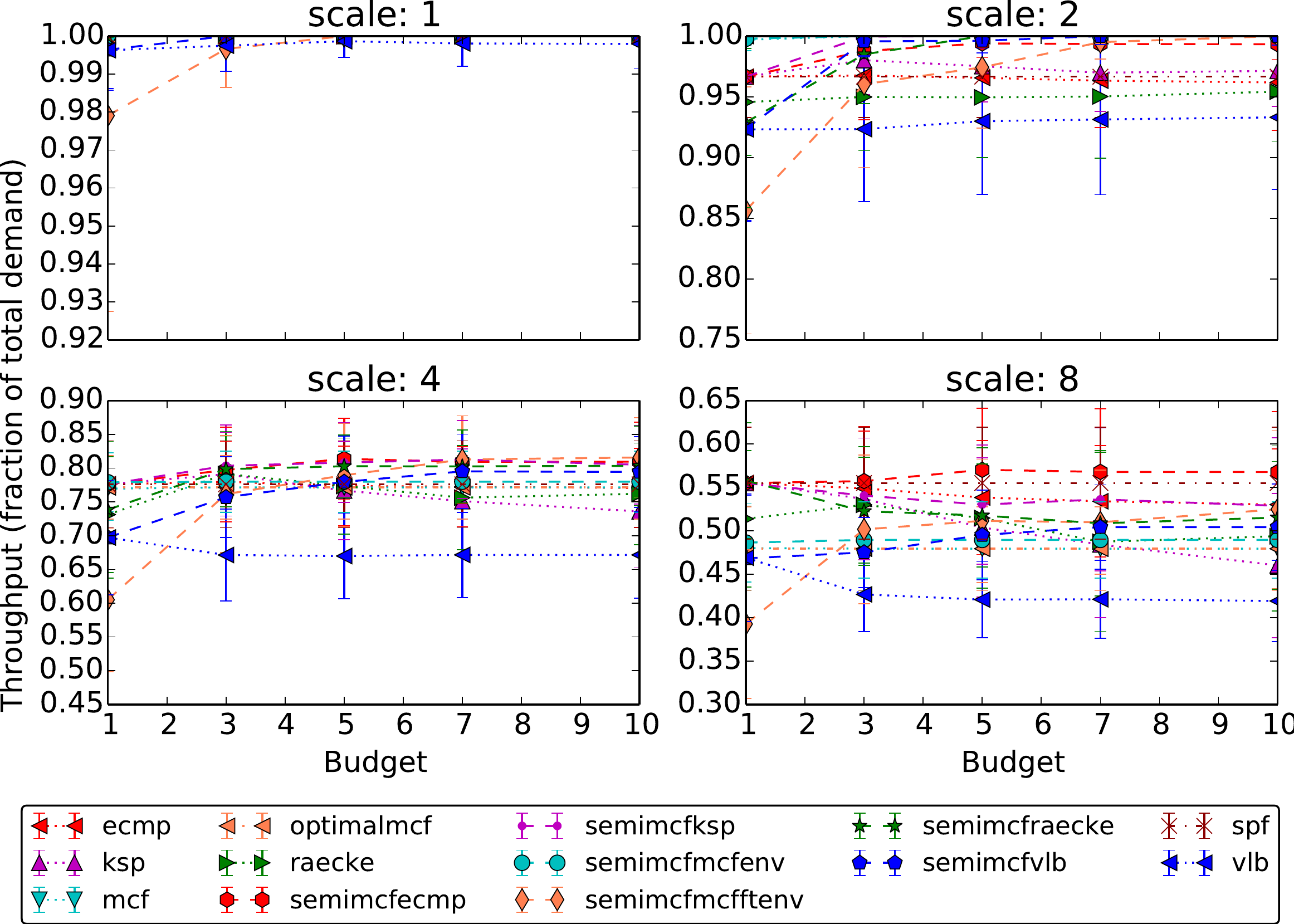}}
\fi
\caption{Throughput as path budget increases.}
\label{fig:budget}
\end{figure}


\textbf{What are the operational overheads for different algorithms?}
We also measure various operational overheads such as churn, number of
paths used and time to solve. Figure~\ref{fig:overheads} show these
overheads averaged over all the topologies. \emph{With respect to solver
time, solving MCF over the entire set of possible paths is 2 orders of
magnitude slower than SemiMcf based schemes where the base set of
paths for each pair of source and destination is a small
constant.}  LP-based solutions to MCF are very sensitive to slight
changes in input. As a result with slight variations in traffic
matrices over time, the set of paths used by MCF can vary greatly,
resulting in high churn. When performing local recovery for failures,
the base path set for all routing schemes, except OptimalMCF, didn't
change. Thus, they don't incur any recovery churn. OptimalMCF solved
MCF using the updated topology and thus has non-zero recovery
churn. When measuring the number of paths used, we set budget to
3. Thus, most routing schemes ended up using their entire
limit. However, algorithms like SPF and ECMP could use only a limited
number of paths well within the budget. In our experiments, the number
of paths chosen by MCF was close to 1 per pair on average.

\iftechreport
\begin{figure}[!ht]
\centerline{\includegraphics[width=\columnwidth]{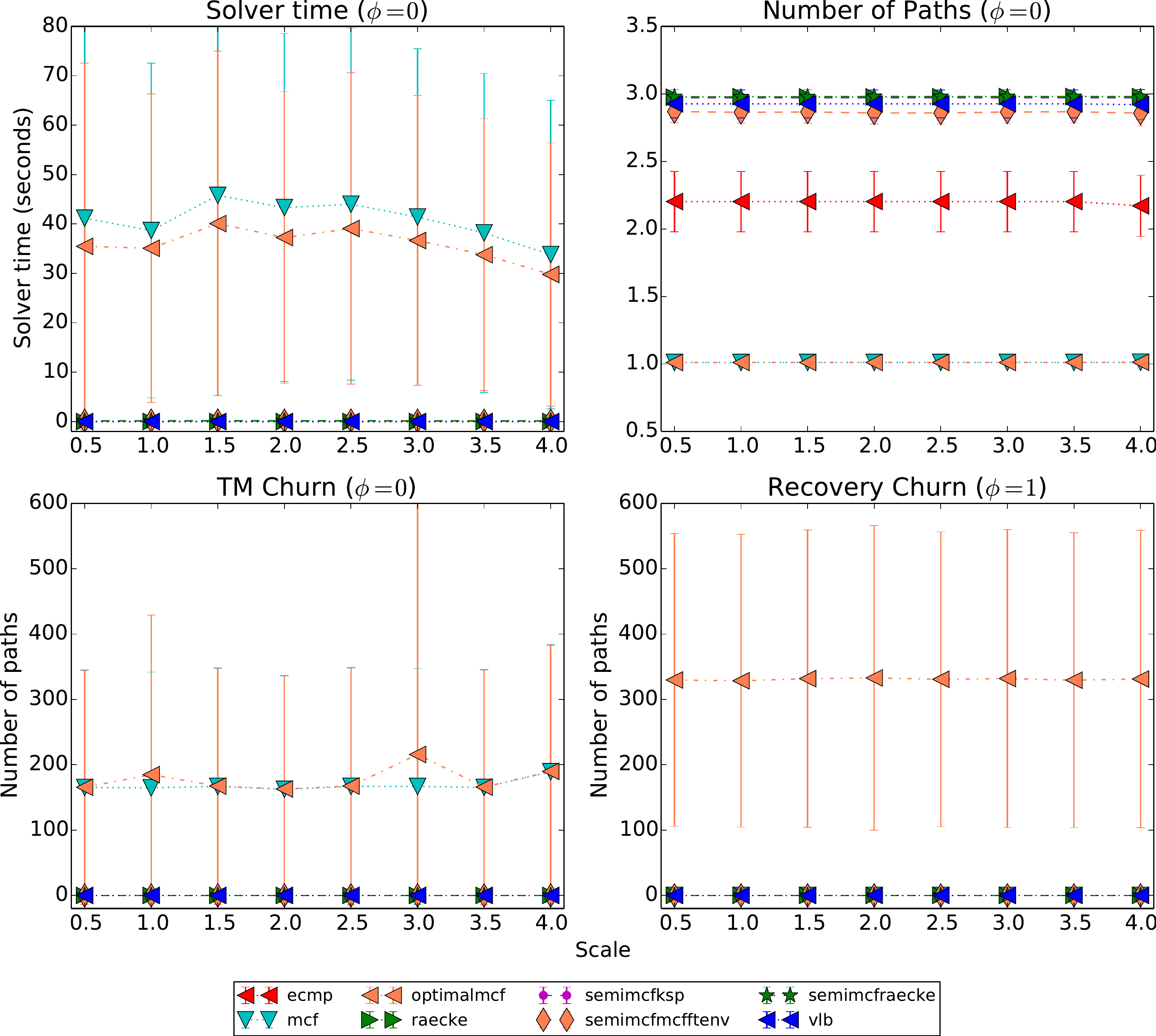}}
\else
\begin{figure}[t]
\centerline{\includegraphics[width=\columnwidth]{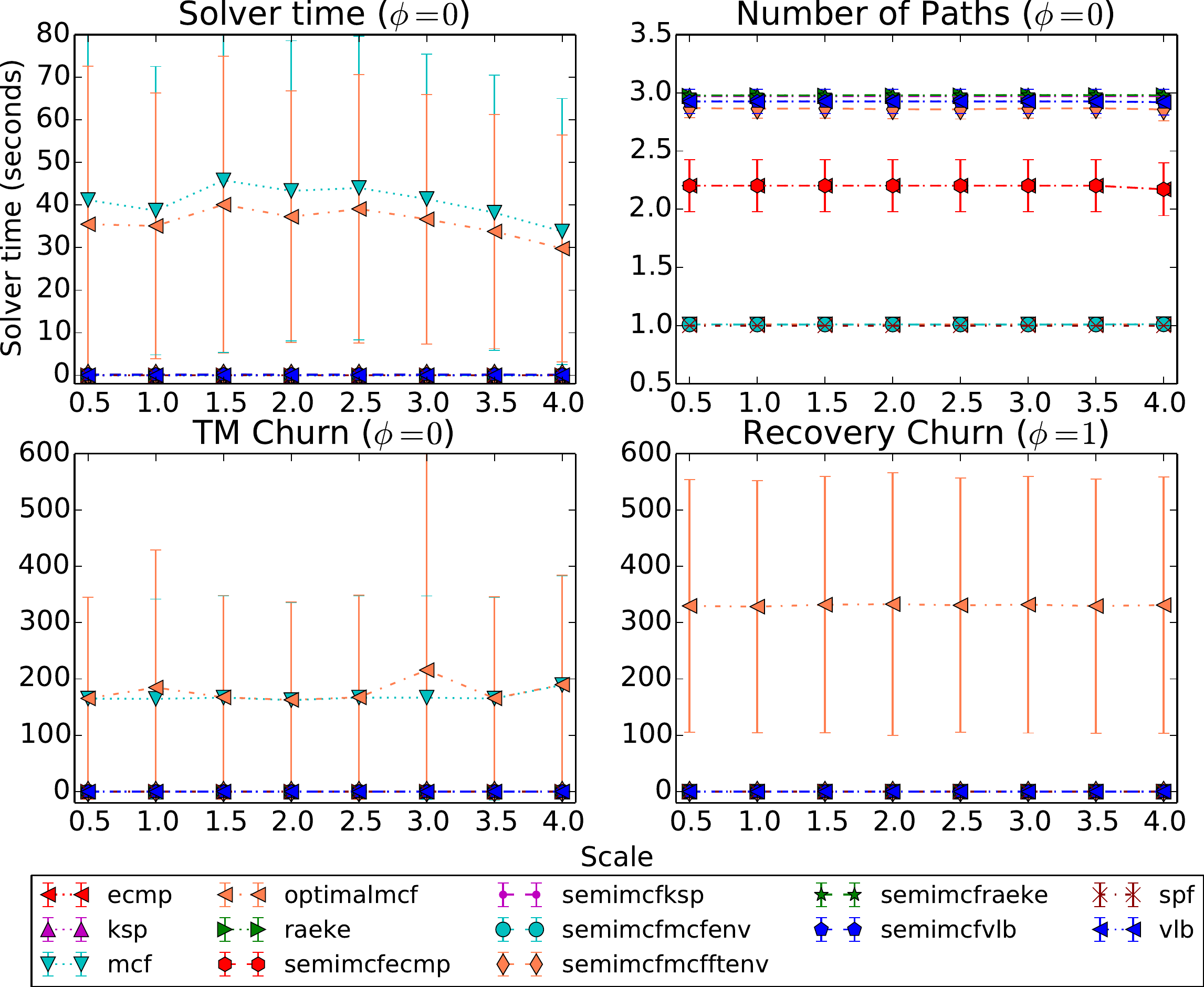}}
\fi
\caption{Various operational overheads associated with different routing schemes.}
\label{fig:overheads}
\end{figure}

\clearpage

\section{Related Work}
\label{sec:related}


There has been considerable interest in traffic engineering for
wide-area networks, which we've already surveyed in
Section~\ref{sec:background}. More broadly, there have been a number
of recent systems for traffic engineering in the data
center~\cite{alfares10,
ballani11,perry14,popa12,jeyakumar13,popa13,shieh10b}. While clearly
related, our work is focused on wide area networks, which face
different requirements.

Several systems have proposed using randomized routing techniques.
RouteBricks~\cite{argyraki08, dobrescu09} extends the basic Valiant
load balancing algorithm to guard against packet-reordering, and to
constant-degree graph topologies. Zhang-Shen and McKeown~\cite{shen05,
shen08} have proposed VLB for use with backbone networks, and show
that the networks can guarantee 100\% throughput for any traffic
matrix, even in the event of link failures.  Kodialam et
al.~\cite{kodialam09,kodialam08} proposed the use of oblivious routing
in IP backbone networks. Applegate and Cohen~\cite{applegate03} showed
that, in practice, oblivious routing performed better than the
worst-case theory predictions.

The discovery of oblivious routing schemes with polylogarithmic
congestion in general networks by R\"{a}cke in 2002~\cite{racke02}
sparked significant interest in the theory community. The performance
of R\"{a}cke's original scheme was improved in a series of
papers~\cite{BKR03,HHR03} culminating in the optimal $O(\log n)$
scheme~\cite{racke08} used by our system.  For any network topology,
it is known how to compute an oblivious routing scheme with the
optimal worst-case congestion ratio, in polynomial time, using the
ellipsoid method~\cite{azar2003} or interior point
methods~\cite{applegate03}. Oblivious routing schemes with a polylogarithmic,
rather than logarithmic, congestion ratio can be computed in almost
linear time~\cite{rst14}.  Hajiaghayi et al.~\cite{hajiaghayi2005}
derived improved bounds in models where the traffic matrix has
independent random entries rather than worst-case entries.

\section{Conclusion}
\label{sec:conclusion}

Operators of wide-area networks face competing requirements when
implementing a traffic engineering strategy.  On the one hand, to
improve operational efficiency, they seek to improve network
utilization by distributing load evenly amongst capacitated links. On
the other hand, to mitigate the overhead of network management, they
seek to minimize the frequency and complexity of state changes on
devices. The semi-oblivious traffic engineering (SOTE) scheme
presented in this paper provides a balanced solution to the competing
requirements of traffic engineering, while at the same time,
simplifying the management infrastructure.

\label{ConcPage}

{\small
\bibliographystyle{abbrv}
\balance
\bibliography{main}
}

\end{document}